\documentclass[aps,prb,reprint,amssymb]{revtex4-2}

\usepackage{amsmath,amssymb,color,braket}
\usepackage{bm}
\usepackage{hyperref}
\usepackage{graphicx}

\begin{document}

\title{Spin functional renormalization group for dimerized quantum spin systems }

\author{Andreas R\"{u}ckriegel}
\affiliation{Institut f\"{u}r Theoretische Physik, Universit\"{a}t Frankfurt,  Max-von-Laue Stra{\ss}e 1, 60438 Frankfurt, Germany}

\author{Jonas Arnold}
\affiliation{Institut f\"{u}r Theoretische Physik, Universit\"{a}t Frankfurt,  Max-von-Laue Stra{\ss}e 1, 60438 Frankfurt, Germany}

\author{Raphael Goll}
\affiliation{Institut f\"{u}r Theoretische Physik, Universit\"{a}t Frankfurt,  Max-von-Laue Stra{\ss}e 1, 60438 Frankfurt, Germany}

\author{Peter Kopietz}
\affiliation{Institut f\"{u}r Theoretische Physik, Universit\"{a}t Frankfurt,  Max-von-Laue Stra{\ss}e 1, 60438 Frankfurt, Germany}

\date{June 14, 2022}

%
%
%
%
%
%
%
%
%
%
%
%
%

\begin{abstract}
We investigate dimerized quantum spin systems using 
the spin functional renormalization group approach proposed by 
Krieg and Kopietz [Phys.~Rev.~B {\bf 99}, 060403(R) (2019)] which
directly focuses on the physical spin correlation functions and
avoids the representation of the spins in terms
of fermionic or bosonic  auxiliary operators.
Starting from decoupled dimers as initial condition 
for the renormalization group flow equations,
we obtain the spectrum of the triplet excitations as well as the magnetization
in the quantum paramagnetic, ferromagnetic, and thermally disordered phases
at all temperatures.
Moreover,
we compute the full phase diagram of a weakly coupled  dimerized spin system
in three dimensions,
including the correct mean field critical exponents
at the two quantum critical points.
\end{abstract}

\maketitle

\section{Introduction}

%
%
%
%
%

Dimerized spin systems are quantum Heisenberg magnets 
where a dominant antiferromagnetic interaction between two 
neighboring spins, which form a dimer unit,
enforces a singlet ground state 
at small magnetic fields~\cite{Ruegg2003,Giamarchi2008,Zapf2014}.
The system is then a quantum paramagnet which does not exhibit 
long-range magnetic order.
Low-energy excitations above this ground state can be viewed as 
gapped bosonic quasiparticles
whose density can be controlled by the external magnetic field $H$.
As the magnetic field increases,
these excitations undergo two separate Bose-Einstein condensation (BEC) 
quantum phase transitions: in $D=3$ spatial dimensions
the first transition
is characterized by the emergence of
$XY$ antiferromagnetic order at a critical field $H_{c1}$.
When the magnetic field is further increased, there is a second transition
at $H = H_{c2}$ to a fully polarized ferromagnetic state.
These BEC quantum phase transitions and the associated intrinsic quantum fluctuations of dimerized quantum spin systems 
have attracted considerable interest in recent years, 
both experimentally and theoretically~\cite{Nikuni2000,Cavadini2000,Oosawa2001,Ruegg2003,Sherman2003,Jaime2004,Matsumoto2004,
Nohadani2004,Ruegg2005,Giamarchi2008,Sebastian2006,Ruegg2007,Batista2007,Zapf2014,Zhou2020}.
Prominent materials which have been shown to be well described
by quantum dimer models include 
${\rm TlCuCl}_3$ \cite{Nikuni2000,Oosawa2001,Cavadini2001,Cavadini2002,Matsumoto2002,Ruegg2003,Sherman2003,
Matsumoto2004,Sirker2005,Ruegg2005,Zhou2020}, 
${\rm KCuCl}_{3}$ \cite{Cavadini1999,Cavadini2000,Cavadini2002}, and 
${\rm BaCuSi}_2 {\rm O}_6$ \cite{Sasago1997,Jaime2004,Sebastian2006,Ruegg2007,Batista2007},
among many others \cite{Zapf2014}.

The basic features of the phase diagram of dimerized quantum spin systems
have already been revealed in $1970$ via 
mean-field theory \cite{Tachiki1970}.
However, in several respects the mean-field results compare poorly with 
experiments.
For example,
mean-field theory fails to reproduce the power law behavior of the critical temperature
that is expected for the BEC quantum phase transition
and has been observed experimentally~\cite{Nikuni2000,Oosawa2001,Sherman2003,Nohadani2004,Sebastian2006,Zapf2014,Zhou2020}.
Another drawback of this mean-field approach is that
it does not directly deal with the physical spin operators
but with auxiliary spin-$1/2$ operators which capture only
the two lowest states of the dimer.
While this reduction of the Hilbert space allows for a simple
mean-field description of the quantum paramagnetic phase, 
it breaks down at elevated temperatures,
where the higher states cannot be neglected,
as well as for small magnetic fields,
where the Zeeman splitting between the excited states becomes small.
A more sophisticated method to study  dimerized spin systems 
is based on the representation of the spin operators in terms of suitably defined auxiliary bosons~\cite{Nikuni2000,Matsumoto2002,Jaime2004,Matsumoto2004,Sirker2005,
Ruegg2005,Batista2007,Zapf2014,Zhou2020}.
However, this strategy also has some disadvantages:
first of all,
the mapping to auxiliary Bose operators obscures the direct connection to 
the physical spin operators which tends to obscure the 
physical interpretation of the results.
Moreover,
the Hilbert space of the Bose operators contains unphysical states
which should be eliminated by means of some projection procedure,
such as an infinite on-site repulsion.
At elevated temperatures,
one furthermore has to account for the thermal reweighting of the dimer states 
by an appropriate ansatz \cite{Ruegg2005}.

In this work, we study dimerized quantum spin systems
using the functional renormalization group (FRG) approach
to quantum spin systems recently developed in 
Refs.~[\onlinecite{Krieg2019,Tarasevych2018,Goll2019,Goll2020,Tarasevych2021,Tarasevych2022}].
Our spin FRG approach generalizes and extends earlier work by
Machado and Dupuis \cite{Machado10} who developed a 
lattice FRG  group  approach for classical spin systems.
Although later the lattice  FRG was also used to
study bosonic quantum lattice models \cite{Rancon11a,Rancon11b, Rancon12a,Rancon12b,Rancon14},
the direct application of this method  
to quantum spin systems was not possible due to some technical difficulties related to the
existence of the average effective action of quantum Heisenberg models with spin-rotational invariance. In Refs.~[\onlinecite{Krieg2019,Tarasevych2018,Goll2019,Goll2020,Tarasevych2021,Tarasevych2022}] we have developed several strategies to avoid these technical difficulties.
In contrast to methods based on auxiliary bosons \cite{Nikuni2000,Matsumoto2002,Jaime2004,Matsumoto2004,Sirker2005,
Ruegg2005,Batista2007,Zapf2014,Zhou2020}, our spin FRG
directly manipulates the physical spin correlation functions, 
thus circumventing all issues associated with the expression of
quantum spins in terms of  bosonic or fermionic auxiliary degrees of freedom.
We show in particular that a straightforward truncation of the spin FRG flow equations
yields good results for the excitation spectrum and thermodynamics
of weakly coupled dimers
outside of the antiferromagnetic $XY$ phase at all temperatures and magnetic fields,
including the critical fields where the system exhibits BEC
quantum phase transitions.
We also obtain the correct (mean field) critical exponents 
at the BEC quantum critical points in dimension $D=3$,
and compute corrections to the lower critical field due to quantum fluctuations.

The rest of this work is organized as follows:
In Sec.~\ref{sec:dimer} we define a model Hamiltonian 
for a  dimerized quantum spin system and discuss its phase diagram
qualitatively. 
In Sec.~\ref{sec:flow},
we then formulate the spin FRG for our dimerized quantum spin systems,
develop a truncation strategy for the flow equations,
and present our FRG results for the mode spectrum and the phase diagram.
Finally,  in Sec.~\ref{sec:discussion}
we conclude with  a summary of our main results and an outlook 
on future research directions.
In two  appendices 
we give additional technical details: in Appendix~\ref{app:dimer}
we explicitly give the imaginary-time-ordered spin correlation functions
of a single dimer involving up to four spins which are needed to calculate the
initial values of the vertices in our spin FRG flow equations.
Appendix~\ref{app:SFRG} contains a brief description of our spin FRG formalism.

\section{Dimerized quantum spin systems}

\label{sec:dimer}

%
%

\begin{figure}[tb]
\centering
\includegraphics[width=\linewidth]{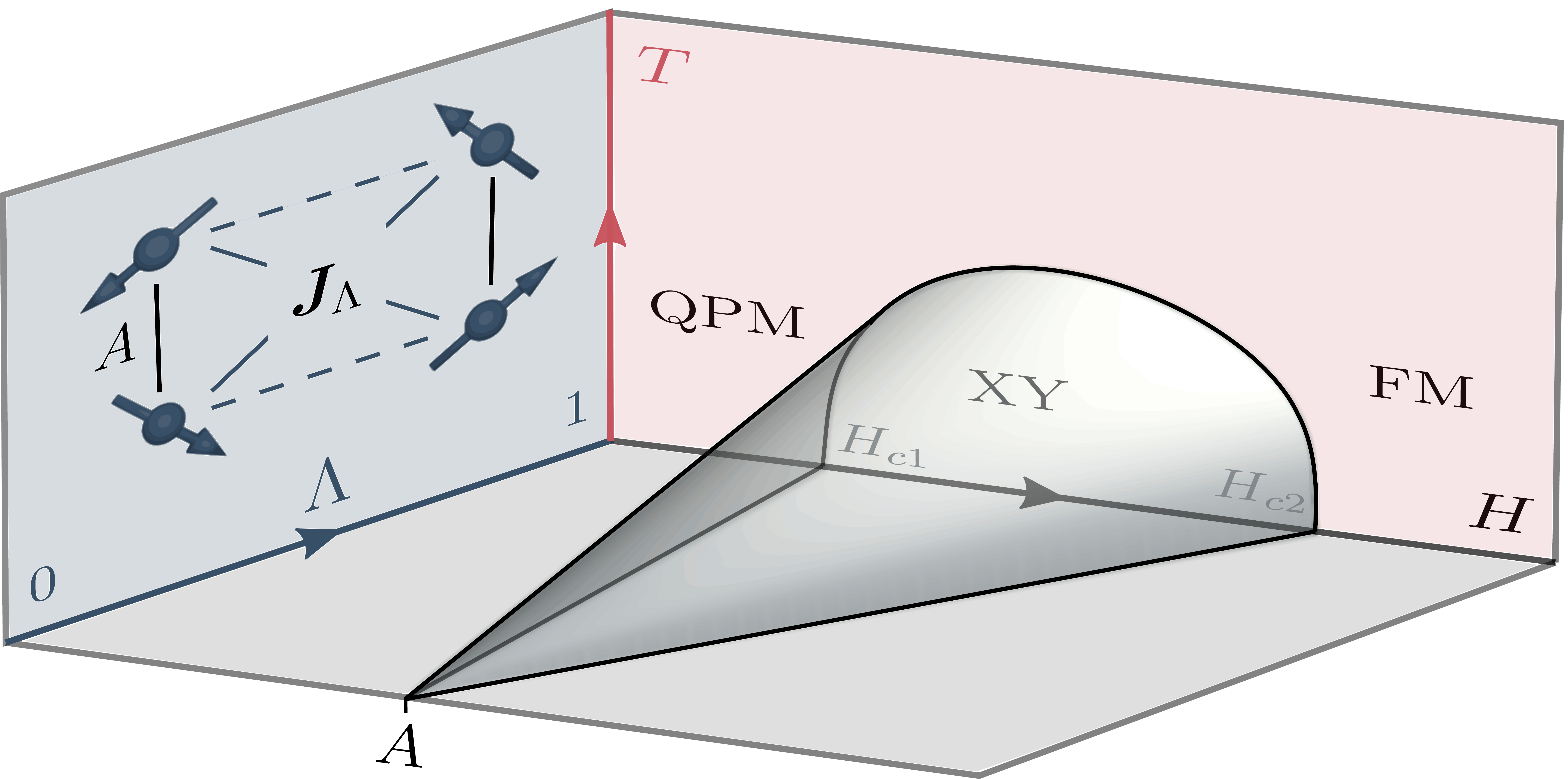}
\caption{
Schematic depiction of the phase diagram of a dimerized spin system
as a function of the FRG deformation parameter $\Lambda$.
At the beginning of the flow (where $\Lambda = 0$ and ${\bf{J}}_{\Lambda =0} =0$)
the dimers are completely decoupled.
Then the $T=0$ phase diagram consists only of the quantum paramagnetic (QPM) 
and the ferromagnetic (FM) phase,
separated by a quantum critical point at $H=A$.
When the inter-dimer exchange couplings ${\bf J}_\Lambda$ are turned on
with increasing deformation parameter $\Lambda$,
this quantum critical point grows into an additional phase
exhibiting antiferromagnetic $XY$ ordering in $D=3$ dimensions.
The new quantum critical points at the critical fields $H_{c1}$ and $H_{c2}$ respectively
separate the QPM and FM phases from the $XY$ phase.
Inset:
Visualization of a pair of spin dimers,
with intra-dimer exchange $A$ and (deformed) inter-dimer exchange couplings ${\bf J}_\Lambda$.
}
\label{fig1}
\end{figure}

The essential physics of  dimerized quantum spin systems
is  described by the following quantum Heisenberg spin Hamiltonian,
\begin{align}
{\cal H} 
= {} &
\frac{ 1 }{ 2 } \sum_{ i j } \sum_{ n  m = 1 }^2 \left( 
J_{ i j, n m }^\bot  \bm{s}_{ i , n }^\bot \cdot \bm{s}_{ j , m }^\bot + J_{ i j, n m }^\parallel s_{ i , n }^z s_{ j , m }^z 
\right)
\nonumber\\
&
+ A \sum_i \bm{s}_{ i , 1 } \cdot \bm{s}_{ i , 2 }
- H \sum_{ i } \sum_{ n = 1 }^2 s_{ i , n }^z .
\label{eq:H_spins}
\end{align}
Here, 
$\bm{s}_{ i , n } = ( s_{ i , n }^x , s_{ i , n }^y , s_{ i , n }^z ) = ( \bm{s}_{ i , n }^\bot , s_{ i , n }^z )$ 
are spin-$1/2$ operators associated with  dimer $i = 1 , \ldots , N$
at magnetic site $n = 1 , 2$.
The dimers are coupled antiferromagnetically via the inter-dimer exchange couplings 
$ J_{ i j, n m }^\alpha > 0 $ (where $\alpha = \bot , \parallel$),
which are assumed to be small compared to the antiferromagnetic intra-dimer exchange $A > 0$.
Lastly,
$H$ is the Zeeman energy associated with an external magnetic field in $z$ direction.
Such a system is illustrated schematically in the inset of Fig.~\ref{fig1}.

Introducing the total and staggered dimer spin operators as
\begin{subequations} \label{eq:ST}
\begin{align}
\bm{S}_i = {} & \bm{s}_{ i , 1 } + \bm{s}_{ i , 2 } , \\
\bm{T}_i = {} & \bm{s}_{ i , 1 } - \bm{s}_{ i , 2 } , 
\end{align}
\end{subequations}
respectively,
we can rewrite the Hamiltonian \eqref{eq:H_spins} as
\begin{equation}
{\cal H} = {\cal H}_0 + {\cal V} + {\rm const} ,
\end{equation}
where
\begin{subequations} \label{eq:H_0}
\begin{align}
{\cal H}_0 = {} & \sum_{ i = 1 }^N h_i , \\ 
h_i = {} & \frac{ A }{ 2 } \bm{S}_i^2 - H S_i^z 
\label{eq:h_dimer}
\end{align}
\end{subequations}
is the Hamiltonian of a collection of $N$ decoupled dimers,
and
\begin{align} \label{eq:V}
{\cal V} 
= {} & \frac{ 1 }{ 2 } \sum_{ i j } \left( 
J_{ T , i j }^\bot \bm{T}_i^\bot \cdot \bm{T}_j^\bot + J_{ T , i j }^\parallel T_i^z T_j^z \right.
\nonumber\\
& \phantom{ \frac{ 1 }{ 2 } \sum_{ i j } }
+ \left. 
J_{ S , i j }^\bot \bm{S}_i^\bot \cdot \bm{S}_j^\bot + J_{ S , i j }^\parallel S_i^z S_j^z
\right) 
\end{align}
describes the exchange interactions between the dimers.
Note that by writing down the exchange Hamiltonian \eqref{eq:V},
we have assumed for simplicity that the two 
magnetic sites of a given dimer are equivalent,
such that $J_{ i j, 1 1 }^\alpha = J_{ i j, 2 2 }^\alpha$.
The relevant exchange couplings for $\alpha = \bot , \parallel$ are then given by
\begin{subequations}
\begin{align}
J_{ S , i j }^\alpha = {} & 
\frac{ 1 }{ 2 } \left( J_{ i j, 1 1 }^\alpha + J_{ i j, 1 2 }^\alpha \right) , \\
J_{ T , i j }^\alpha = {} & 
\frac{ 1 }{ 2 } \left( J_{ i j, 1 1 }^\alpha - J_{ i j, 1 2 }^\alpha \right) .
\end{align}
\end{subequations}
For inequivalent magnetic sites,
there is an additional $\bm{S}_i \cdot \bm{T}_j$ exchange coupling.
However, 
we will see below that for weakly coupled dimers at low energies,
this additional coupling does not give rise to relevant interaction processes 
because the dynamics of the total and staggered spin operators are well separated in energy.

Before proceeding further,
it is instructive to consider the Hamiltonian \eqref{eq:h_dimer} 
of a single dimer in more detail.
Its eigenstates are given by the singlet state
 \begin{equation} 
 \ket{ s }_i = \frac{ 1 }{ \sqrt{ 2 } } \left(  
\ket{ \uparrow \downarrow }_i - \ket{ \downarrow \uparrow }_i
\right), 
 \end{equation}
and the three Zeeman-split triplet states,
 \begin{subequations}
 \begin{align}
 \ket{ t + }_i & =  \ket{ \uparrow \uparrow }_i , 
 \\
 \ket{ t 0 }_i & =  \frac{ 1 }{ \sqrt{ 2 } } \left(  
\ket{ \uparrow \downarrow }_i + \ket{ \downarrow \uparrow }_i
\right) ,
 \\
 \ket{ t - }_i & = \ket{ \downarrow \downarrow }_i . 
 \end{align}
 \end{subequations}
The corresponding eigenenergies are
\begin{subequations} \label{eq:dimer_energies}
\begin{align}
E^{s} = {} & 0 , \\
E^{+} = {} & A - H , \\
E^{0} = {} & A , \\
E^{-} = {} & A + H .
\end{align}
\end{subequations}
From these energies
it is obvious  that at the magnetic field $H=A$
an isolated  dimer exhibits at zero temperature a quantum phase transition from
the singlet state, which is a quantum paramagnet,
to the fully polarized $+$ triplet state.
At finite temperatures $T = 1/  \beta > 0$
the four dimer states are thermally occupied, 
with Boltzmann factors
\begin{subequations} \label{eq:Boltzmann}
\begin{align}
p^{ 0 } = {} & p^{ s } e^{ - \beta E^{0} } , \\
p^{ \pm } = {} & p^{ s } e^{ - \beta E^{\pm } } , \\
p^{ s } = {} & \frac{ 1 }{ 
1 + e^{ - \beta E^{+} } + e^{ - \beta E^{0} } + e^{ - \beta E^{-} }
} .
\end{align}
\end{subequations}
These Boltzmann factors and the associated eigenenergies 
fully determine all correlation functions of the single dimer.
It turns out that with our truncation of the FRG flow equations we need
time-ordered single-dimer correlation functions involving up to four powers of the spin operators. In spite of the simplicity of the single-dimer
Hamiltonian, these correlation functions have a highly non-trivial frequency dependence. We summarize the relevant expressions in Appendix~\ref{app:dimer}.

The inter-dimer exchange ${\cal{V}}$ in
Eq.~\eqref{eq:V} has a two-fold effect on the properties of a single dimer:
firstly, 
it endows the eigenenergies \eqref{eq:dimer_energies} with a dispersion,
and secondly,
it enables interaction between the different eigenstates of the isolated dimer.
This also leads to the emergence of a new phase 
at the quantum critical point $H=A$ of the isolated dimer:
in dimension $D=3$,
this phase exhibits XY antiferromagnetic long-range order as indicated 
in Fig.~\ref{fig1},
while  in $D=2$ and $D=1$ it corresponds to a Berezinskii-Kosterlitz-Thouless 
or a Luttinger liquid phase, respectively \cite{Zapf2014}.

\section{FRG Flow equations for dimerized quantum spin systems}

\label{sec:flow}

%
%
%
%

\subsection{Spin FRG}

\label{sec:SFRG}

The spin FRG approach proposed in Ref.~[\onlinecite{Krieg2019}]
and further developed in Refs.~[\onlinecite{Tarasevych2018,Goll2019,Goll2020,Tarasevych2021,Tarasevych2022}] 
is based on a formally exact renormalization group flow equation for the generating functional of connected spin correlation functions.
As such, 
it does not require projecting the physical spin operators onto auxiliary bosons or fermions with restricted Hilbert spaces. 
In fact, 
the spin FRG combines the old spin-diagram technique 
developed by Vaks, Larkin and Pikin \cite{Vaks1967a,Vaks1967b,Izyumov1988} 
with modern FRG methods \cite{Wetterich1993,Berges2002,Pawlowski2007,Kopietz2010,Metzner2012,Dupuis2021}.
It turns out that the spin FRG flow equation is formally equivalent to the bosonic Wetterich equation \cite{Krieg2019},
which allows us to utilize the established diagrammatic FRG techniques for bosons \cite{Kopietz2010}, 
thus avoiding the more complicated diagrammatic rules
of the spin diagram technique \cite{Vaks1967a,Vaks1967b,Izyumov1988} .
The non-trivial $SU(2)$ algebra of the spin operators is taken into account via  
non-trivial initial conditions for the flow equations.

To set up the spin FRG in the context of our dimerized spin system \eqref{eq:H_spins},
we replace the inter-dimer exchange couplings $J_{ a , i j }^\alpha$ (where $a = S , T$)
by deformed couplings $J_{ \Lambda , a , i j }^\alpha$.
Here, the continuous parameter $\Lambda \in [ 0 , 1 ]$ plays the role of the flowing cutoff in the FRG.
The deformed couplings $J_{ \Lambda , a , i j }^\alpha$ are to be chosen such 
that $J_{ \Lambda = 1 , a , i j }^\alpha = J_{ a , i j }^\alpha$,
while for $\Lambda =0$ the model should be simple enough to allow for a controlled solution. 
For a dimerized spin system,
a natural choice is  $J_{ \Lambda = 0 , a , i j }^\alpha = 0$.
In this case,
the Hamiltonian at the initial scale is given by the Hamiltonian (\ref{eq:H_0}) of decoupled dimers,
which is exactly solvable and already contains information on both the quantum disordered state at low magnetic fields and temperatures
as well as on the quantum phase transition to the ferromagnetic state at elevated magnetic fields;
see Appendix~\ref{app:dimer}.
The phase diagram resulting from this flow is schematically depicted in Fig.~\ref{fig1}.

In the following,
we will consider the FRG flow of a special hybrid functional $\Gamma_\Lambda [ \varphi ]$
which generates irreducible vertices with the following properties:
the vertices should be  (a) one-line irreducible with respect to all three components of 
the staggered spin propagators; (b) one-line irreducible with respect to the two 
transverse components of the total spin propagators; and (c) the vertices should be
interaction-irreducible with respect to cutting 
a longitudinal inter-dimer interaction between the total spins.
The explicit construction of a functional $\Gamma_\Lambda [ \varphi ]$
with these properties has been discussed in Refs.~[\onlinecite{Goll2019,Tarasevych2021}]
and is reviewed in Appendix~\ref{app:SFRG}.
At imaginary time $\tau$,
the superfield $\varphi_{ a , i }^\alpha ( \tau )$ then corresponds for $a = S$ and
$\alpha = x,y$
to the local transverse magnetization,
for $a=S$ and $\alpha = z$ to the
the fluctuating part of the local inter-dimer longitudinal exchange field,
and for $a = T$ and $\alpha = x,y,z$ 
to the three components the local staggered spin.
The six components of the superfield $\varphi  = ( \varphi_{ a , i }^\alpha ( \tau ) )  $, where the flavor index $a = S, T $ refers to the total and the
staggered spin of a given dimer, are then explicitly given by
 \begin{equation}
  ( \varphi_{ a , i }^\alpha ( \tau ) ) = 
 \left( \begin{array}{c}  \varphi_{S, i}^x ( \tau ) \\ \varphi_{S, i}^y ( \tau ) \\ \varphi_{S, i}^z ( \tau ) \\
 \varphi_{T, i}^x ( \tau ) \\ \varphi_{T, i}^y ( \tau ) \\ \varphi_{T, i}^z ( \tau ) \end{array}
 \right) =
\left( \begin{array}{c}  \langle S_{ i}^x ( \tau ) \rangle \\ \langle S_{ i}^y ( \tau ) \rangle \\ 
 \varphi_{ i} ( \tau ) \\
 \langle T_{ i}^x ( \tau ) \rangle \\ \langle T_{ i}^y ( \tau ) \rangle \\ \langle 
T_{ i}^z ( \tau ) \rangle \end{array}
 \right) ,
 \end{equation}
where the longitudinal total-spin exchange field $\varphi_i ( \tau )$ is defined 
in Appendix~\ref{app:SFRG}, see also Ref.~[\onlinecite{Goll2019}].
The different treatment of the longitudinal total spin of a dimer
is necessitated by the $U(1)$ spin-rotational symmetry around the $z$ axis
of the Hamiltonian (\ref{eq:H_0}) of decoupled dimers at the initial scale,
which implies that the longitudinal magnetization field has no 
dynamics at $\Lambda = 0$ when the coupling between the dimers is switched off \cite{Goll2019}.
For details on the derivation of this functional and the associated flow equations,
we refer to Appendix~\ref{app:SFRG} and to
Refs.~[\onlinecite{Krieg2019,Goll2019}].

For a given value of the deformation parameter $\Lambda$,
the vertex expansion of our hybrid  generating functional is of the form 
\begin{widetext}
\begin{align}
\Gamma_\Lambda [ \varphi ]
= {} &
\beta N f_\Lambda
+ \int_K \sum_{ a = S , T } \left[
\Gamma_{ \Lambda, a a }^{ + - } ( -K, K ) \varphi_a^- ( - K ) \varphi_a^+ ( K ) 
+ \frac{ 1 }{ 2 ! } \Gamma_{ \Lambda, aa }^{ z z } ( -K , K ) 
\varphi_a^z ( - K ) \varphi_a^z ( K )
\right]
\nonumber\\
&
+ \int_{ K_1 K_2 K_3 } \delta\left( K_1 + K_2 + K_3 \right) \left[ 
\sum_{a = S , T } \Gamma_{ \Lambda , aa S }^{ + - z } ( K_1 , K_2 , K_3 )
\varphi_a^- ( K_1 ) \varphi_a^+ ( K_2 ) \varphi_S^z ( K_3 ) 
\vphantom{ \frac{ 1 }{ 3 ! } }
\right.
\nonumber\\
& \phantom{ + }
+ \Gamma_{ \Lambda, T S T }^{ + - z } ( K_1 , K_2 , K_3 ) 
\varphi_T^- ( K_1 ) \varphi_S^+ ( K_2 ) \varphi_T^z ( K_3 )
+ \Gamma_{ \Lambda, S T T }^{ + - z } ( K_1 , K_2 , K_3 ) 
\varphi_S^- ( K_1 ) \varphi_T^+ ( K_2 ) \varphi_T^z ( K_3 )
\nonumber\\
& \phantom{ + } \left.  \vphantom{ \sum_{s = S , T } }
+ \frac{ 1 }{ 2 ! } \Gamma_{ \Lambda, T T S }^{ z z z } ( K_1 , K_2 , K_3 ) 
\varphi_T^z ( K_1 ) \varphi_T^z ( K_2 ) \varphi_S^z ( K_3 )
+ \frac{ 1 }{ 3 ! } \Gamma_{ \Lambda, S S S }^{ z z z } ( K_1 , K_2 , K_3 ) 
\varphi_S^z ( K_1 ) \varphi_S^z ( K_2 ) \varphi_S^z ( K_3 )
\right]
\nonumber\\
&
+ {\cal O} \left( \varphi^4 \right) ,
\label{eq:Gamma_expansion}
\end{align}
\end{widetext}
where
$K = ( \bm{k} , i\omega )$ is a collective label for momentum $\bm{k}$ and Matsubara frequency $i\omega$; the  corresponding integration and delta symbols are defined as
$\int_K = ( \beta N )^{ - 1 } \sum_{ \bm{k} , i \omega }$ and
$\delta ( K ) = \beta N \delta_{ \bm{k} , 0 } \delta_{ \omega , 0 }$, respectively.
Here $\varphi_a^\pm = ( \varphi_a^x \pm i \varphi_a^y ) / \sqrt{2} $ denote 
the spherical transverse field components and
the field-independent contribution $f_\Lambda$ can be identified with the flowing free energy per dimer.
Note that in writing down the vertex expansion \eqref{eq:Gamma_expansion},
we already took into account two symmetries of the Hamiltonian \eqref{eq:H_spins}:
the global $U(1)$ spin-rotational symmetry around the $z$ axis that corresponds to spin conservation,
as well as the invariance under exchange of the two magnetic sites;
i.e., $\bm{s}_{ i , 1} \leftrightarrow \bm{s}_{ i , 2 }$ or $\bm{T}_i \leftrightarrow - \bm{T}_i$.
The former implies that only vertex functions with the same number of $+$ and $-$ labels are finite,
while the latter requires all vertices to have an even number of $T$ labels. 
In doing so,
we have of course neglected the possibility of spontaneous symmetry breaking
that is necessary to describe the $XY$ ordered phase of dimerized spin systems in dimension $D=3$ \cite{Nikuni2000,Giamarchi2008,Zapf2014}.
Although it is possible to extend our spin FRG approach to include also the XY ordering, 
this is beyond the scope of the present work.
The 2-point vertex functions determine the flowing spin propagators via
\begin{subequations} \label{eq:propagators_general}
\begin{align}
G_{ \Lambda , a }^\bot ( K ) 
= {} &
\frac{ 1 }{ 
\Gamma_{ \Lambda, aa }^{ + - } ( -K, K )
+ J_{ \Lambda , a , \bm{k} }^\bot - J_{ a , \bm{k} }^\bot
} , 
\label{eq:propagators_perp_general}
\\
G_{ \Lambda , T }^\parallel ( K )
= {} &
\frac{ 1 }{ 
\Gamma_{ \Lambda, T T }^{ z z } ( -K, K )
+ J_{ \Lambda , T , \bm{k} }^\parallel - J_{ T , \bm{k} }^\parallel
} , 
\label{eq:propagator_Tz_general}
\\
G_{ \Lambda , S }^\parallel ( K )
= {} &
\frac{ \Pi_\Lambda ( K ) }{ 1 + J_{ \Lambda , S , \bm{k} }^\parallel \Pi_\Lambda ( K ) } ,
\end{align}
\end{subequations}
where 
$ J_{ \Lambda , a , \bm{k} }^\alpha $ is the Fourier transform of $J_{ \Lambda , a, i j  }^\alpha$
and 
\begin{equation}
\Pi_\Lambda ( K ) = 
- \Gamma_{ \Lambda , S S }^{  z z } ( -K, K ) - \frac{ 1 }{ J_{ S , \bm{k} }^\parallel }
\end{equation}
is the interaction-irreducible longitudinal spin susceptibility~\cite{Goll2019}.
The decoupled-dimer initial conditions for these propagators are listed in 
Eqs.~\eqref{eq:2-point_dimer}.
Note that the propagators \eqref{eq:propagators_general} are the $2$-spin correlation functions,
which directly determine quantities of experimental interest like the dynamical spin structure factor.

\begin{figure}[tb]
\centering
\includegraphics[width=\linewidth]{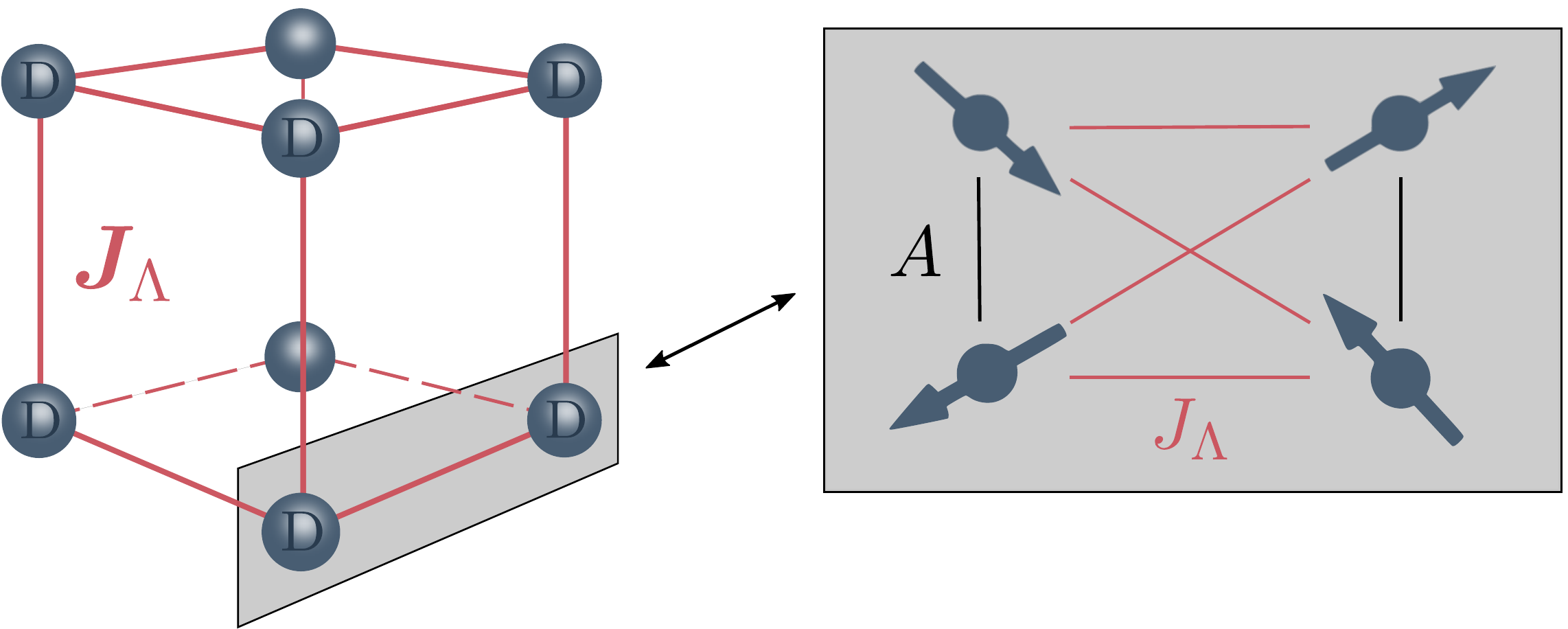}
\caption{
Illustration of a simple cubic lattice of dimers (D) in three dimensions.
The dimers interact via (deformed) isotropic inter-dimer exchange couplings ${\bf J}_\Lambda$.
The two spins that form a given dimer interact with each other via the intra-dimer exchange $A$.
}
\label{fig:simple_cubic}
\end{figure}

For the explicit calculations in this work,
we assume that the dimers form a simple cubic lattice in three dimensions
with lattice constant $a$,
and that all inter-dimer exchange interactions are isotropic.
This setup is illustrated in Fig.~\ref{fig:simple_cubic}.
Then we can write the flowing exchange couplings as 
\begin{equation}
J^\alpha_{ \Lambda, a , \bm{k} } 
= J^\alpha_{ a , \bm{k} = 0 } \gamma_{ \Lambda ,  \bm{k} } ,
\end{equation}
with 
$ \gamma^\alpha_{ \Lambda = 0 , \bm{k} } = 0 $
and
$ \gamma^\alpha_{ \Lambda = 1 , \bm{k} } = \gamma^\alpha_{ \bm{k} } $.
Here,
\begin{equation}
\gamma_{ \bm{k} }
= \frac{ 1 }{ 3 } \left[ 
\cos \left( k^x a \right) + \cos \left( k^y a \right) + \cos \left( k^z a \right) 
\right]
\end{equation}
is the nearest neighbor form factor,
which satisfies
$ -1 \le \gamma_{ \bm{k} } \le 1 $.
For the deformation scheme,
we use a Litim regulator \cite{Litim2001}, 
given by
\begin{align}
\label{eq:Litim}
\gamma_{ \Lambda , \bm{k} } 
&=
{\rm sgn} \left( \gamma_{ \bm{k} } \right)
\left.
\begin{cases}
\Lambda , & \Lambda < \left| \gamma_{ \bm{k} } \right| \\
\left| \gamma_{ \bm{k} } \right| , & \Lambda \ge \left| \gamma_{ \bm{k} } \right|
\end{cases}
\right\}
 \nonumber
\\
&=
\gamma_{ \bm{k} } -
{\rm sgn} \left( \gamma_{ \bm{k} } \right)
\left( \left| \gamma_{ \bm{k} } \right| - \Lambda \right)
\Theta \left( \left| \gamma_{ \bm{k} } \right| - \Lambda \right) ,
\end{align}
such that
\begin{equation}
\partial_\Lambda \gamma_{ \Lambda , \bm{k} } =
{\rm sgn} \left( \gamma_{ \bm{k} } \right)
\Theta \left( \left| \gamma_{ \bm{k} } \right| - \Lambda \right) .
\end{equation}
Physically,
the flow then corresponds to increasing the bandwidth of all exchange interactions 
from $0$ to the final value $2$.
This deformation scheme has the advantage that  
closed loop integrations that frequently appear in the spin FRG flow equations
can be performed analytically as follows,
\begin{equation} \label{eq:Litim_loop}
\int_{ \bm{k} } \left( \partial_\Lambda \gamma_{ \Lambda , \bm{k} } \right)
F_\Lambda \left( \gamma_{ \Lambda , \bm{k} } \right)
=
n ( \Lambda ) \left[ F_\Lambda ( \Lambda ) - F_\Lambda ( - \Lambda ) \right] .
\end{equation}
Here, 
$\int_{ \bm{k} } = N^{ - 1 } \sum_{ \bm{k} }$,
$F_\Lambda \left( \gamma_{ \Lambda , \bm{k} } \right)$ 
is an arbitrary function of the deformed exchange coupling,
and
\begin{equation}
n ( \Lambda ) = \int_\Lambda^1 d \epsilon\, \nu( \epsilon ) 
\end{equation}
is an effective number of states between dimensionless energies $\Lambda$ and $1$,
with the density of states
\begin{equation}
\nu ( \epsilon ) = 
\int_{ \bm{k} } \delta\left( \gamma_{ \bm{k} } - \epsilon \right)
\end{equation}
of the exchange interaction.
The two functions $n(\Lambda)$ and $\nu(\epsilon)$ 
can be computed once for a given structure factor 
and can then be used subsequently in all flow equations. 
We stress at this point that the spin FRG approach 
is applicable to any lattice structure.
Different lattices only modify the density of states $\nu(\epsilon)$,
without altering the form of the spin FRG flow equations.
Here, 
we consider an isotropic simple cubic lattice
both for simplicity and to facilitate a direct comparison to quantum Monte Carlo results in Sec.~\ref{sec:quantum}.
For any unfrustrated lattice,
we furthermore expect qualitatively similar results.

\subsection{Tadpole resummation}

\label{sec:tadpole}

\begin{figure}[tb]
\centering
\includegraphics[width=\linewidth]{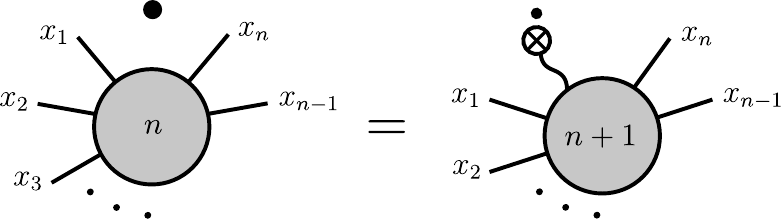}
\caption{
Diagrammatic representation of the tadpole flow equations \eqref{eq:tadpole_flow}.
The filled circles with $n$ external legs represent $n$-point vertices.
The labels of each leg are collected into $x_i = ( \alpha_i s_i K_i )$.
The crossed circle with a wavy line represents the flowing exchange field $\phi_\Lambda$.
Dots above one of these elements signify a scale derivative $\partial_\Lambda$.
}
\label{fig:tadpole}
\end{figure}

At finite magnetic field $H$ and temperature $T$,
the system  possesses a finite magnetization $M$.
In addition to the $n$-point vertex functions, we should then  
also keep track of the renormalization group flow of the longitudinal inter-dimer exchange field
\begin{equation} \label{eq:exchange_field}
\phi_\Lambda = - J_{ \Lambda , S , \bm{k} = 0 }^\parallel M_\Lambda 
\end{equation}
which is determined by the flowing magnetization $M_\Lambda$.
Neglecting for the moment all terms in the  flow equations for the irreducible vertices 
involving loop integrations, we find that the flow of the  exchange field
generates the following infinite hierarchy  of flow equations for the $n$-point vertices
with $n \geq 2$ external legs,
%
%
%
\begin{align} 
&
\partial_\Lambda 
\Gamma_{ \Lambda, a_1 \ldots a_n }^{ \alpha_1 \ldots \alpha_n } ( K_1 , \ldots , K_n ) 
\nonumber\\
= {} &
\Gamma_{ \Lambda, a_1 \ldots a_n S }^{ \alpha_1 \ldots \alpha_n z } ( K_1 , \ldots , K_n , 0 ) 
\partial_\Lambda \phi_\Lambda .
\label{eq:tadpole_flow}
\end{align}
Graphically, these flow equations correspond to the tadpole diagrams 
displayed in Fig.~\ref{fig:tadpole}.
Integrating the tadpole flow equation \eqref{eq:tadpole_flow}
from $0$ to $\Lambda$ and iterating we obtain the explicit solution
\begin{align}
&
\Gamma_{ \Lambda, a_1 \ldots a_n }^{ \alpha_1 \ldots \alpha_n } ( K_1 , \ldots , K_n ) 
\nonumber\\
= {} &
\sum_{ m = 0 }^\infty \frac{ \phi_\Lambda^m }{ m ! }
\Gamma_{ 0 , a_1 \ldots a_n \underbrace{ \scriptstyle S \ldots S }_m }^{ \alpha_1 \ldots \alpha_n \overbrace{ \scriptstyle z \ldots z }^m } 
( K_1 , \ldots , K_n , \overbrace{ 0 , \ldots , 0 }^m ) 
.
\label{eq:tadpole_sol_series}
\end{align}
We now note that by definition of
our hybrid functional $\Gamma_{\Lambda} [ \varphi ]$
given in Eq.~(\ref{eq:Gamma_functional}) of Appendix~\ref{app:SFRG}
(see also Ref.~[\onlinecite{Goll2019}]) 
the initial vertex functions appearing 
on the right-hand side of the solution \eqref{eq:tadpole_sol_series}
can be  related to lower-order vertex functions
by taking derivatives with respect to the magnetic field $H$,
\begin{align}
&
\Gamma_{ 0 , a_1 \ldots a_n \underbrace{ \scriptstyle S \ldots S }_m }^{ \alpha_1 \ldots \alpha_n \overbrace{ \scriptstyle z \ldots z }^m } 
( K_1 , \ldots , K_n , \overbrace{ 0 , \ldots , 0 }^m ) 
\nonumber\\
= {} &
\partial_H^m \Gamma_{ 0 , a_1 \ldots a_n  }^{ \alpha_1 \ldots \alpha_n } ( K_1 , \ldots , K_n ) .
\end{align}
Therefore we can write the solution \eqref{eq:tadpole_sol_series} of the 
tadpole flow equation \eqref{eq:tadpole_flow} as
%
\begin{align}
&
\Gamma_{ \Lambda, a_1 \ldots a_n }^{ \alpha_1 \ldots \alpha_n } ( K_1 , \ldots , K_n ) 
\nonumber\\
= {} &
\sum_{ m = 0 }^\infty \frac{ \phi_\Lambda^m }{ m ! }
\partial_H^m \Gamma_{ 0 , a_1 \ldots a_n  }^{ \alpha_1 \ldots \alpha_n } ( K_1 , \ldots , K_n )
 \nonumber
\\
= {} &
\left.
\Gamma_{ 0 , a_1 \ldots a_n  }^{ \alpha_1 \ldots \alpha_n } ( K_1 , \ldots , K_n ) 
\right|_{ H \to H + \phi_\Lambda } .
\label{eq:GammaallH}
\end{align}
%
Hence, 
the resummation of the tadpole diagrams shown in Fig.~\ref{fig:tadpole} alone 
simply yields the mean field shift of the magnetic field
\begin{equation} \label{eq:H_Lambda}
H \to H + \phi_\Lambda = H - J_{ \Lambda , S , \bm{k} = 0 }^\parallel M_\Lambda 
\end{equation}
in \emph{all} vertex functions.
Properly including this shift during the flow proves to be crucial to obtain physically meaningful results in the following calculations.

In a simple truncation where only these tadpole diagrams are taken into account,
the flowing staggered spin propagators are given by
\begin{subequations} \label{eq:G_T_tadpole}
\begin{align}
G_{ \Lambda , T }^\bot ( K )
= {} &
\frac{ M_0 \left( i\omega + A - H - \phi_\Lambda \right) - 2 m_0 A 
}{
\left( E^{ + }_{ \Lambda , \bm{k} } + i\omega \right)
\left( E^{ - }_{ \Lambda , \bm{k} } - i\omega \right)
} ,
\label{eq:G_T_transverse_tadpole}
\\
G_{ \Lambda , T }^\parallel ( K )
= {} &
\frac{  2 \left( p^{ s } - p^ { 0 } \right) A 
}{
\left( E^{ 0 }_{ \Lambda , \bm{k} } + i\omega \right)
\left( E^{ 0 }_{ \Lambda , \bm{k} } - i\omega \right)
} ,
\end{align}
\end{subequations}
where 
 \begin{equation}
 M_0 = p^{ + } - p^{ - } 
 \end{equation}
is the magnetic moment  of an isolated dimer and
 \begin{equation} 
 m_0 = p^{ + } - p^{ s } ,
 \end{equation} 
where the Boltzmann factors should be evaluated 
at the flowing magnetic field $H+\phi_\Lambda$.
Note that $ m_0 $ corresponds to the difference in occupation 
of the two lowest energy states of an isolated dimer
and thus measures whether the dimer is closer to the disordered singlet state
or the fully polarized $+$ triplet state.
The flowing dispersion relations of the three triplet states are in this approximation 
given by
\begin{subequations} \label{eq:E_triplet}
\begin{align}
E^{ \pm }_{ \Lambda , \bm{k} }
= {} & 
\sqrt{ 
\left( A + \frac{ M_0 }{ 2 } J_{ \Lambda , T , \bm{k} }^\bot \right)^2 
- 2 m_0 A J_{ \Lambda , T , \bm{k} }^\bot
}
\nonumber\\
&
\mp \left(
\frac{ M_0 }{ 2 } J_{ \Lambda , T , \bm{k} }^\bot + H + \phi_\Lambda
\right)
, \\
E^{ 0 }_{ \Lambda , \bm{k} }
= {} & 
\sqrt{ 
A^2 + 2 \left( p^{ s } - p^ { 0 } \right) A J_{ \Lambda , T , \bm{k} }^\parallel
} 
.
\end{align}
\end{subequations}

\begin{figure}[tb]
\centering
\includegraphics[width=\linewidth]{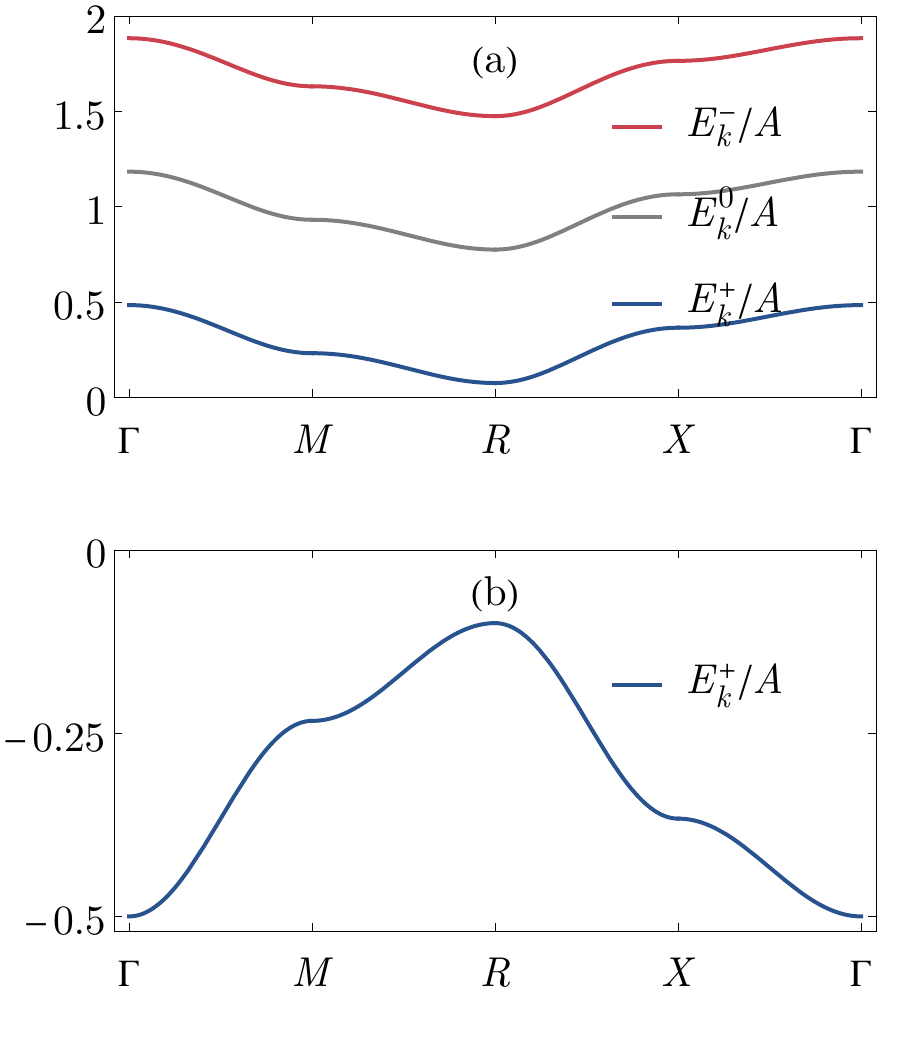}
\caption{
Zero-temperature triplet dispersions for a simple cubic lattice 
with inter-dimer exchange couplings
$J_{ T , \bm{k} = 0 }^\bot = J_{ T , \bm{k} = 0 }^\parallel = J_0 = 0.2 \, A$,
for
(a) $H = 0.7 \, A$ in the quantum paramagnetic phase, and
(b) $H = 1.5 \, A$ in the ferromagnetic phase.
}
\label{fig:dispersion}
\end{figure}

Because  $M_0$ and $m_0$ are complicated functions of temperature 
and (flowing) magnetic field determined by the exact dimer correlation functions,
the properties of the triplet modes \eqref{eq:E_triplet} 
vary considerably over the phase diagram.
In particular,  for $T \ll A $ and $ H < H_{ c 1 } $ 
the system is in the quantum paramagnetic regime 
($M \approx 0$); in this case  
we also have
$M_0 \approx 0 \approx p^{ 0 }$ and $m_0 \approx - p^{ s } \approx -1$.
Then the triplet dispersions \eqref{eq:E_triplet}  reduce at the end of the flow to 
\begin{subequations} 
 \label{eq:E_para}
\begin{align}
E_{ \bm{k} }^{ \pm } 
\approx {} & \sqrt{ A^2 + 2 A J_{ T , \bm{k} }^\bot } \mp H , \\
E^{ 0 }_{ \bm{k} }
\approx {} & 
\sqrt{ A^2 + 2 A J_{ T , \bm{k} }^\parallel } .
\end{align}
\end{subequations}
in agreement with calculations for dimerized quantum spin systems based
on the random-phase approximation \cite{Sasago1997,Cavadini1999,Cavadini2000}. 
These dispersions are shown in Fig.~\ref{fig:dispersion} (a).
It is then easy to see 
that the gap of the lowest ($+$) triplet dispersion 
vanishes at the magnetic field
\begin{equation} \label{eq:hc1}
H_{ c 1 } = \sqrt{ A^2 + 2 A {\rm min}_{\bm{k}} J_{ T , \bm{k} }^\bot } < A ,
\end{equation}
which gives a first approximation 
for the lower quantum critical field of the dimerized spin system.
This value of the quantum critical field of course still lacks corrections
due to quantum fluctuations \cite{Nohadani2004} 
described by  the loop integrations neglected 
in Eq.~\eqref{eq:GammaallH}.
In Sec.~\ref{sec:quantum} we will explicitly calculate the effect of quantum fluctuations
on the critical fields.

Another regime where our general expressions \eqref{eq:E_para} simplify
is the ferromagnetic phase where $ M \approx 1 $.
Assuming   $T \ll A$ and $ H > H_{ c 2 } $ we then have $M_0 \approx 1 \approx m_0$ 
and $p^{ s } \approx 0 \approx p^{ 0 }$.
Then the two high-energy triplet modes disappear completely 
from the propagators \eqref{eq:G_T_tadpole},
so that at the end of the flow we obtain 
a spin-wave like dispersion for the remaining low-energy mode, 
\begin{equation} \label{eq:E_ferro}
 E_{ \bm{k} }^{ + } \approx 
 A -H +  J_{ S , \bm{k} = 0 }^\parallel  -  J_{ T , \bm{k} }^\bot
;
\end{equation}
see Fig.~\ref{fig:dispersion} (b).
The gap of this mode vanishes at the upper critical field
\begin{equation}
H_{ c 2 } = A + J_{ S , \bm{k} = 0 }^\parallel - {\rm min}_{\bm{k}} J_{ T , \bm{k} }^\bot > A .
\end{equation}
Note that at the quantum critical points themselves,
the $+$ dispersion in Eq.~\eqref{eq:E_para}  and Eq.~\eqref{eq:E_ferro}
is at long-wavelengths  quadratic in $\bm{k}$ 
for generic inter-dimer exchange couplings.
Therefore the dynamical critical exponent is $z=2$,
as expected for BEC quantum critical points \cite{Zapf2014}.
When approaching either of the quantum critical fields $ i = 1 , 2 $
at zero temperature, 
the gap of the $+$ mode furthermore vanishes as
$ | H - H_{ c i } | $,
implying the correlation length critical exponent $\nu=1/z=1/2$ \cite{Zapf2014}.

Next, consider the regime of
elevated temperatures (above the antiferromagnetic dome) and for 
flowing magnetic fields $ H + \phi_\Lambda \approx A $,
that is, 
in the vicinity of the critical field of the isolated dimer.
Then we may approximate $m_0 \approx 0$ and
the flowing dispersion $E_{ \Lambda , \bm{k} }^{ + }$
of the lowest triplet state vanishes as well.
However,
since the energy $E_{ \Lambda , \bm{k} }^{ + }$ also cancels out of the 
associated propagator \eqref{eq:G_T_transverse_tadpole},
this corresponds to a simple level crossing instead of a phase transition.
At this point, 
this  mode changes from a hole-like excitation 
with $\omega < 0$ to a particle-like excitation with $\omega > 0$.

It is important to realize that beyond the simple limits discussed above, 
the flowing triplet dispersions \eqref{eq:E_triplet} contain information on the entire 
phase diagram 
through the Boltzmann factors of the isolated dimer
as well as through the flowing magnetization $M_\Lambda$.
This is similar to the thermal reweighting of the dimer states 
proposed in Ref.~[\onlinecite{Ruegg2005}].
In particular,
the condition that the gap of the lowest $+$ triplet mode vanishes
yields an estimate for the full phase transition curve of the antiferromagnetic dome, 
which  will be discussed in  Sec.~\ref{sec:thermal} and in
Fig.~\ref{fig:M} below.

Lastly, let us also give the flowing transverse total spin propagator and the
longitudinal interaction-irreducible total  spin susceptibility
in the tadpole approximation. The former is given by
\begin{equation}
G_{ \Lambda , S }^\bot ( K ) =
\frac{ M_0 }{ H + \phi_\Lambda + M_0 J_{ \Lambda , S , \bm{k} }^\bot - i\omega } ,
\end{equation}
while the interaction-irreducible total spin susceptibility is
 \begin{equation}
 \Pi_\Lambda (K) = \delta_{ \omega , 0 } \partial_H M_0 ( H + \phi_\Lambda ) .
 \label{eq:Pilontot}
 \end{equation}
In the regimes of interest to us
the effects  of both of these total spin correlation functions 
on  the flow of the other correlation functions can be neglected.
To justify this, we note that
transverse total spin correlation function
is negligible in
 the quantum paramagnetic phase because $M_0 \approx 0$,
and in the ferromagnetic phase at large magnetic fields 
because then it only has a single pole at high energies $\sim H$,
which is not thermally excited.
As far as the longitudinal interaction-irreducible total spin susceptibility 
in Eq.~(\ref{eq:Pilontot}) is concerned,
it is only relevant close to the quantum critical point $H=A$ of the isolated dimer, 
see Appendix~\ref{app:dimer}.
As this point lies deep in the antiferromagnetic dome
where our theory is not applicable in its present form anyway,
we may also neglect it.

To conclude this section,
let us point out that already on the tadpole level,
the spin FRG contains two infinite resummations 
-- self-consistent mean field theory and a random-phase approximation --
of the inter-dimer exchange couplings.
Both of these are inherently non-perturbative.
Thus, 
\emph{any} truncation of the spin FRG flow equations
goes beyond a simple perturbation expansion in the inter-dimer exchange.
While small inter-dimer exchange couplings make the truncation 
of the spin FRG flow equations at a low loop order
a more controlled approximation,
such a truncation can consequently also yield reasonable results
for larger values of these exchange couplings.

\subsection{Thermal fluctuations}

\label{sec:thermal}

\begin{figure}[tb]
\centering
\includegraphics[width=\linewidth]{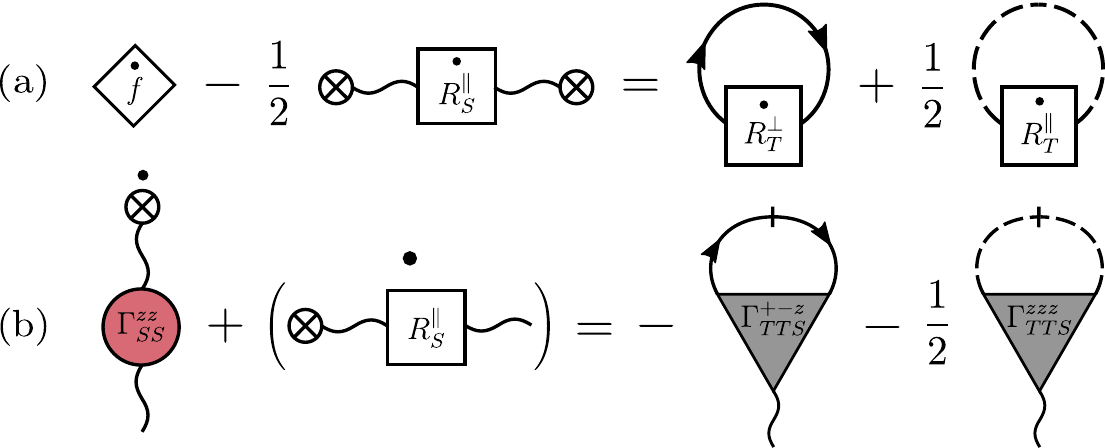}
\caption{
Graphical representation of 
(a) the flow equation \eqref{eq:f_flow} of the free energy and
(b) the flow equation \eqref{eq:phi_flow} of the exchange field.
Solid lines with arrows represent 
the flowing transverse staggered propagator $G_{ \Lambda , T }^\bot ( K )$
and dashed lines the longitudinal staggered propagator $G_{ \Lambda , T }^\parallel ( K )$.
An additional slash marks the corresponding single-scale propagator 
$\dot{ G }_{ \Lambda , T }^\alpha ( K )$.
The rest of the notation is the same as in Fig.~\ref{fig:tadpole}.
}
\label{fig:f_phi}
\end{figure}
At finite temperatures,
thermodynamic quantities such as the magnetization or the specific heat 
are expected to be 
dominated by thermal fluctuations of the dispersive triplet excitations.
To take these properly into account,
we should  include the terms involving loop-integrations on the right-hand 
sides of the corresponding flow equations.
The flow equation for the free energy is
\begin{align} 
&
\partial_\Lambda f_\Lambda 
%
- \frac{ 1 }{ 2 } \phi_\Lambda^2 \partial_\Lambda R_{ \Lambda , S , \bm{k} = 0 }^\parallel
\nonumber\\
= {} &
\int_K 
G_{ \Lambda , T }^\bot ( K ) \partial_\Lambda R_{ \Lambda , T , \bm{k} }^\bot 
+ \frac{ 1 }{ 2 } \int_K 
G_{ \Lambda , T }^\parallel ( K ) \partial_\Lambda R_{ \Lambda , T , \bm{k} }^\parallel ,
\label{eq:f_flow}
\end{align}
while the scale-dependent exchange field  satisfies the flow equation
\begin{align}
&
%
\Gamma_{ \Lambda , S S }^{ z z } ( 0 , 0 ) \partial_\Lambda \phi_\Lambda
+ \partial_\Lambda \left( R_{ \Lambda , S , \bm{k} = 0 }^\parallel \phi_\Lambda \right)
\nonumber\\
= {} & 
-
\int_K 
\Gamma_{ \Lambda , T T S }^{ + - z } ( - K , K , 0 ) \dot{ G }_{ \Lambda , T }^\bot ( K ) 
\nonumber\\
&
-
\frac{ 1 }{ 2 } \int_K
\Gamma_{ \Lambda , T T S }^{ z z z } ( - K , K , 0 )  \dot{ G }_{ \Lambda , T }^\parallel ( K )
.
\label{eq:phi_flow}
\end{align}
Graphical representations of these flow equations are
shown in Fig.~\ref{fig:f_phi}.
Here, 
the staggered single-scale propagators are defined as \cite{Goll2019,Kopietz2010}
\begin{align}
\dot{ G }_{ \Lambda , T  }^\alpha ( K ) 
& = 
\frac{ \partial G_{ \Lambda , T }^\alpha ( K ) 
}{ \partial J_{ \Lambda , T , \bm{k} }^\alpha }
\partial_{\Lambda} { J }_{ \Lambda , T , \bm{k} }^\alpha
 \nonumber
 \\
& = 
- \left[  G_{ \Lambda , T }^\alpha ( K ) \right]^2 
\dot{ J }_{ \Lambda , T , \bm{k} }^\alpha ,
%
%
\end{align}
where
$ \dot{ J }_{ \Lambda , T , \bm{k} }^\alpha = 
\partial_\Lambda J_{ \Lambda , T , \bm{k} }^\alpha $,
and 
\begin{subequations}
\begin{align}
R_{ \Lambda , T , \bm{k} }^\alpha = {} 
&
J_{ \Lambda , T , \bm{k}  }^\alpha
- J_{ T , \bm{k}  }^\alpha , 
\\
R_{ \Lambda , S , \bm{k} }^\parallel = {} 
&
- \frac{ 1 }{ J_{ \Lambda , S , \bm{k}  }^\parallel }
+ \frac{ 1 }{ J_{ S , \bm{k}  }^\parallel }
\end{align}
\end{subequations}
are the staggered spin and 
the longitudinal total spin regulators, see  Refs.[\onlinecite{Goll2019,Goll2020}]
and Appendix~\ref{app:SFRG}.

A general feature of our  spin FRG flow equations for dimerized spin systems,
already apparent in Eqs.~\eqref{eq:f_flow} and \eqref{eq:phi_flow} above,
is that each loop integration is proportional to powers of the flowing inter-dimer exchange couplings $J_{ \Lambda , a , \bm{k} }^\alpha$.
Since we aim to describe dimerized spin systems 
where these couplings are weak compared to the inter-dimer exchange $A$
that we treat exactly via the initial conditions of the spin FRG flow,
we expect that a simple one-loop  truncation of the flow equations 
already yields reasonable results.
For our purpose it is therefore sufficient to
approximate  all vertex functions appearing on the right-hand sides of 
the flow equations 
\eqref{eq:f_flow} and \eqref{eq:phi_flow}
by their tadpole approximations 
discussed in the preceding Sec.~\ref{sec:tadpole}.
That is,
we neglect all loop integrations 
(which give corrections of higher order in the $J_a^\alpha$) 
in their respective flow equations,
but self-consistently replace the magnetic field according to Eq.~\eqref{eq:H_Lambda}
in all dimer correlation functions.
The exact flow equation (\ref{eq:f_flow}) of the free energy then reduces to
\begin{align} 
& 
\partial_\Lambda f_\Lambda 
- \frac{ 1 }{ 2 } M_\Lambda^2 \partial_\Lambda J_{ \Lambda , S , \bm{k} = 0 }^\parallel
\nonumber\\
= {} & 
\frac{ 1 }{ \beta } \int_{ \bm{k} } \dot{ J }^\bot_{ \Lambda , T , \bm{k} } 
\frac{ \partial }{ \partial J^\bot_{ \Lambda , T , \bm{k} } }
\sum_{ r = \pm }
\ln \left( 1 - e^{ r \beta E_{ \Lambda , \bm{k} }^{  r  } } \right)
\nonumber\\
&
+ \frac{ 1 }{ 2 \beta  } \int_{ \bm{k} } 
\dot{ J }^\parallel_{ \Lambda, T , \bm{k} } 
\frac{ \partial }{ \partial J^\parallel_{ \Lambda, T , \bm{k} } }
\sum_{ r = \pm }
\ln \left( 1 - e^{ r \beta E_{ \Lambda , \bm{k} }^{  0  } } \right) ,
\label{eq:f_flow1}
\end{align}
while the flow equation (\ref{eq:phi_flow}) for the exchange field
reduces to the following flow equation for the scale-dependent 
magnetization $M_{\Lambda}$,
\begin{align}
& 
\partial_\Lambda \left[ 
M_\Lambda - M_0 ( H + \phi_\Lambda )
\right]
\nonumber\\
= {} & 
\int_{ \bm{k} } \dot{ J }^\bot_{ \Lambda, T , \bm{k} } 
\frac{ \partial }{ \partial J^\bot_{ \Lambda, T , \bm{k} } }
\sum_{ r = \pm } r 
f_B \left( - r \beta E_{ \Lambda , \bm{k} }^{  r  } \right)
\partial_H E_{ \Lambda , \bm{k} }^{  r  }
\nonumber\\
&
- \frac{ 1 }{ 2 } \int_{ \bm{k} } 
\dot{ J }^\parallel_{ \Lambda, T , \bm{k} } 
\frac{ \partial }{ \partial J^\parallel_{ \Lambda , T , \bm{k} } }
\left[ 1 + 2 f_B \left( \beta E_{ \Lambda , \bm{k} }^{ 0  } \right) \right]
\partial_H E_{ \Lambda , \bm{k} }^{  0  } .
\label{eq:M_flow}
\end{align}
Here,
$ f_B( x ) = 1 / ( e^x - 1 )$ is the Bose function.

\begin{figure}[tb]
\centering
\includegraphics[width=\linewidth]{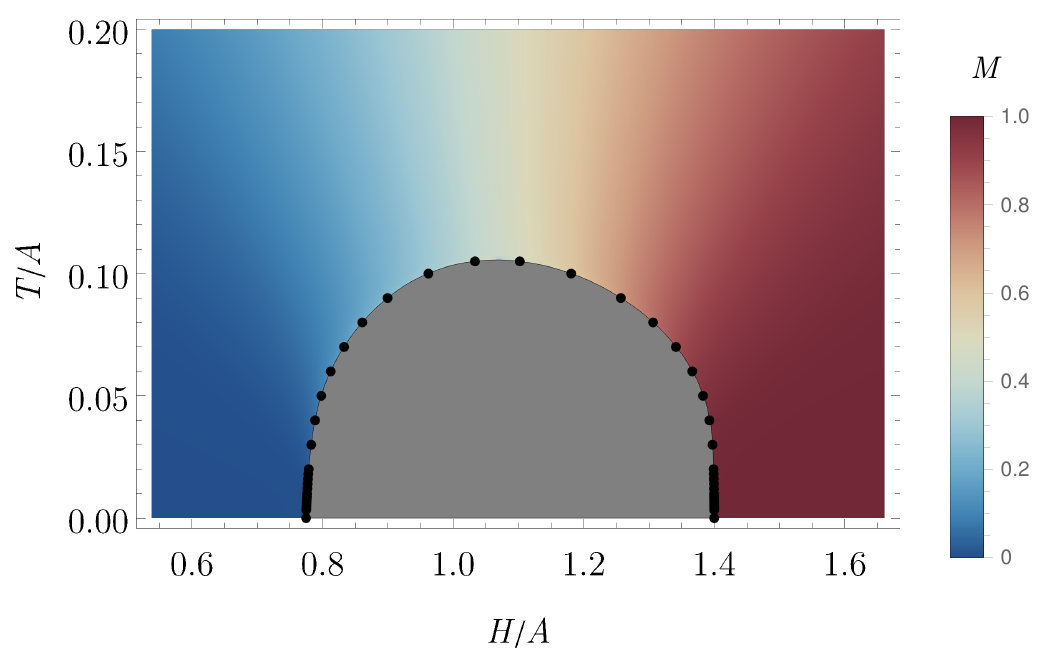}
\caption{
Magnetization of a dimerized spin system in dimension $D=3$
as function of magnetic field $H$ and temperature $T$
for inter-dimer exchange couplings
$J_{ T , \bm{k} = 0 }^\bot = J_{ T , \bm{k} = 0 }^\parallel = J_0 = 0.2 \, A$,
obtained from the numerical solution of the flow equation \eqref{eq:M_flow}.
The black dots are the positions of the extrema of
the susceptibility derivative $ \partial^2 M /  \partial H^2$;
see Fig.~\ref{fig:scaling} (a).
The black line interpolates between these points.
The gray dome enclosed by this line corresponds to the antiferromagnetic $XY$ phase.
There,
our flow equation \eqref{eq:M_flow} is no longer valid
because we do not consider a finite $XY$ order parameter.
}
\label{fig:M}
\end{figure}

\begin{figure*}[tb]
\centering
\includegraphics[width=\linewidth]{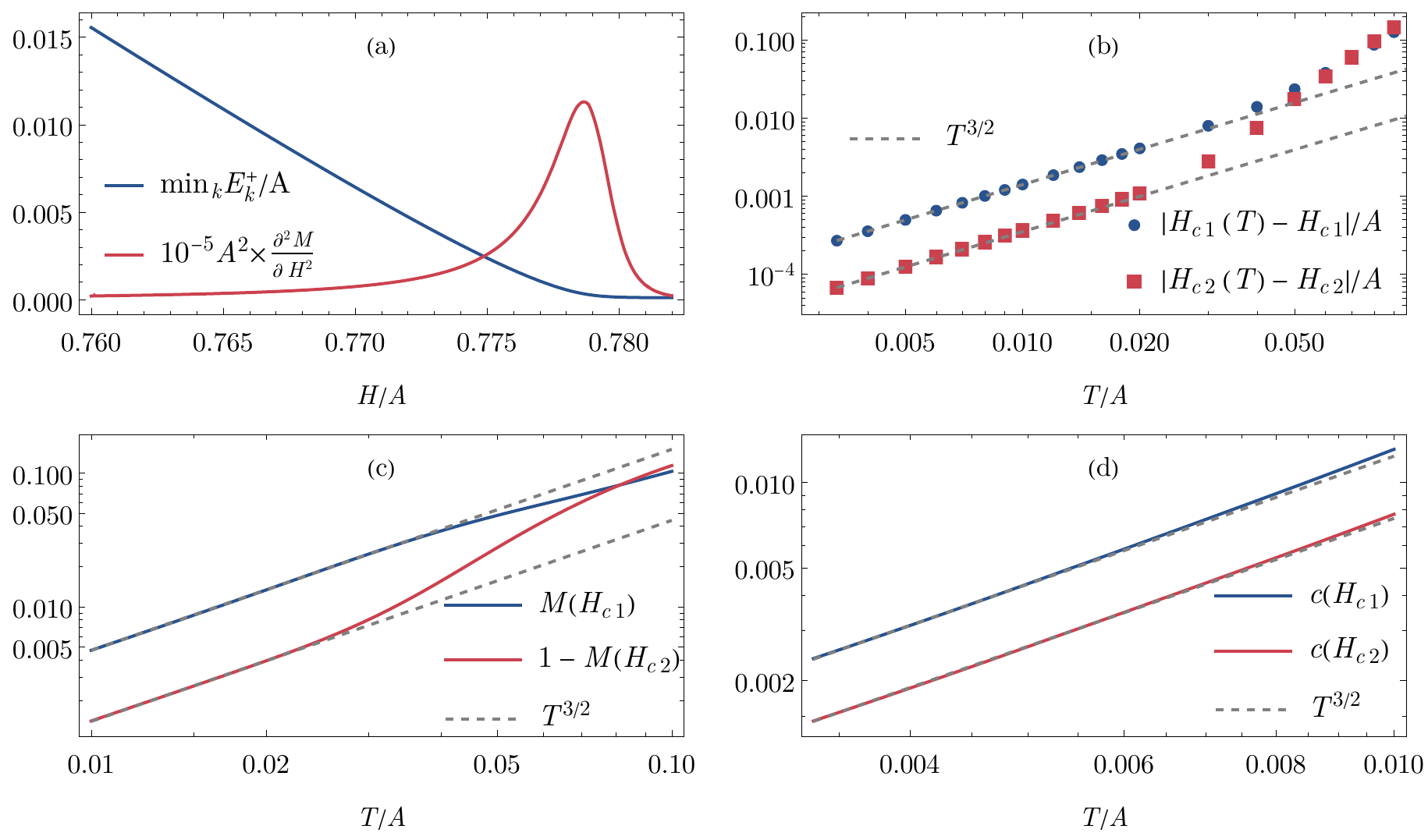}
\caption{
(a) 
Gap ${\rm min}_{\bm{k}} E_{ \bm{k} }^{  +  }$ of the lowest triplet mode
and susceptibility derivative $ \partial^2 M /  \partial H^2$
as function of the magnetic field $H$
at temperature $T=0.02 \, A$.
We also show
log-log plots of the temperature dependence of
(b) the critical fields,
(c) the magnetization at the quantum critical fields, and
(d) the specific heat at the quantum critical fields,
with the expected $T^{3/2}$ power laws for comparison.
All plots are
obtained from the numerical solution of the flow equations \eqref{eq:f_flow1} and \eqref{eq:M_flow},
for inter-dimer exchange couplings
$J_{ T , \bm{k} = 0 }^\bot = J_{ T , \bm{k} = 0 }^\parallel = J_0 = 0.2 \, A$.
}
\label{fig:scaling}
\end{figure*}

The flow equations
\eqref{eq:f_flow1} and \eqref{eq:M_flow} have an simple
interpretation in terms of the
scale derivatives
of the free energies of the bosonic triplet modes
and their magnetic field derivatives,
respectively.
The remaining loop integrations in both of these flow equations 
are of the form given in Eq.~\eqref{eq:Litim_loop}.
Hence, with the Litim regulator \eqref{eq:Litim}
the flow equations \eqref{eq:f_flow1} and \eqref{eq:M_flow}
 reduce to ordinary differential equations 
which  can be straightforwardly integrated numerically.
The resulting magnetization is shown in Fig.~\ref{fig:M}
for inter-dimer exchange couplings
$J_{ T , \bm{k} = 0 }^\bot = J_{ T , \bm{k} = 0 }^\parallel = J_0 = 0.2 \, A$,
as function of magnetic field and temperature.
One clearly sees the quantum paramagnetic phase at small 
and the ferromagnetic phase at large magnetic fields,
as well as the thermally disordered phase at elevated temperatures.
At intermediate fields and low temperatures,
there is additionally the $XY$-ordered dome,
where the flow equation \eqref{eq:M_flow} is no longer applicable.
Numerically,
the boundary of this dome is determined from the critical softening
of the lowest ($+$) triplet mode
and the associated peak in the susceptibility derivative 
$ \partial^2 M /  \partial H^2$, as shown in
Fig.~\ref{fig:scaling}~(a).

With the explicit numerical solution of the flow equations 
\eqref{eq:f_flow1} and \eqref{eq:M_flow} for the free energy 
and the magnetization,
we can furthermore verify the various 
thermodynamic critical exponents that are expected for 
BEC quantum critical points in three dimensions.
At low enough temperatures the relevant power laws are \cite{Zapf2014}
\begin{subequations} \label{eq:exponents}
\begin{align}
\left| H_{ c i } (T) - H_{ c i } \right| \propto {} & T^{ 3 / 2 } , 
\;\;\; i = 1 , 2 , 
\label{eq:phase_boundary}
\\
M ( H_{ c 1 } ) \propto {} & T^{ 3 / 2 } , \\
1 - M ( H_{ c 2 } ) \propto {} & T^{ 3 / 2 } , \\
c ( H_{ c i } ) \propto {} & T^{ 3 / 2 } , 
\;\;\; i = 1 , 2 ,
\end{align}
\end{subequations}
where $ H_{ c i } (T) $ denotes the critical fields 
as function of temperature,
and $c ( H ) = - T \partial^2 f /  \partial T^2$
is the specific heat of the dimerized spin system.
Our results for these  quantities are displayed
in Figs.~\ref{fig:scaling} (b) -- (d) on a $\log$-$\log$ scale,
showing good agreement with the power laws \eqref{eq:exponents}.
It is furthermore apparent
that these asymptotic power laws can only be observed
in a small temperature window \cite{Nohadani2004}.
For example, for the critical field shown in Fig.~\ref{fig:scaling} (b) 
the power law is obeyed up to $T \approx 0.02 \, A = 0.1 J_0$.
Attempting to fit the critical field with a power law in a larger temperature window
yields too large exponents in the range $1.7$ -- $2.1$,
in agreement with previous theoretical predictions and experimental observations 
\cite{Nikuni2000,Oosawa2001,Sherman2003,Nohadani2004,Sirker2005,Zapf2014}.
Ultimately,
this can be traced back to the fact that
the long-wavelength limit of the triplet dispersions
breaks down rather quickly away from the quantum critical point
because of the relative smallness of the inter-dimer exchange
compared to the intra-dimer exchange~\cite{Sherman2003,Nohadani2004,Sirker2005}.

In dimensions $D < 3$,
a qualitatively similar phase diagram can be obtained 
from the solution of the flow equation \eqref{eq:M_flow} for the magnetization.
Unlike in $D=3$ however,
the gap of the lowest triplet mode no longer closes in reduced dimensions at finite temperatures.
This reflects the increased relevance of quantum fluctuations,
which require a more sophisticated truncation of the spin FRG flow equations
that also takes triplet-triplet interactions into account.
We leave this problem for future work.
However,
even in this case an estimate for the critical field 
of the Berezinskii-Kosterlitz-Thouless or Luttinger liquid phase transition in $D=1$ or $2$ 
can be obtained from the peak in the susceptibility derivative,
beyond which the magnetization flows to unphysical values.

\subsection{Quantum fluctuations}

\label{sec:quantum}

\begin{figure}[tb]
\centering
\includegraphics[width=\linewidth]{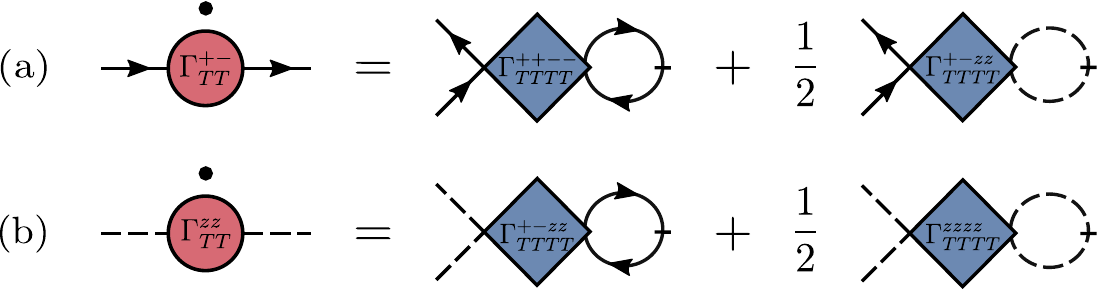}
\caption{
Diagrammatic representation of the flow equations \eqref{eq:T=0_flow}
for the staggered $2$-point vertices.
The meaning of the graphical elements is the same as 
in Figs.~\ref{fig:tadpole} and \ref{fig:f_phi}.
}
\label{fig:gamma2}
\end{figure}

\begin{figure}[tb]
\centering
\includegraphics[width=\linewidth]{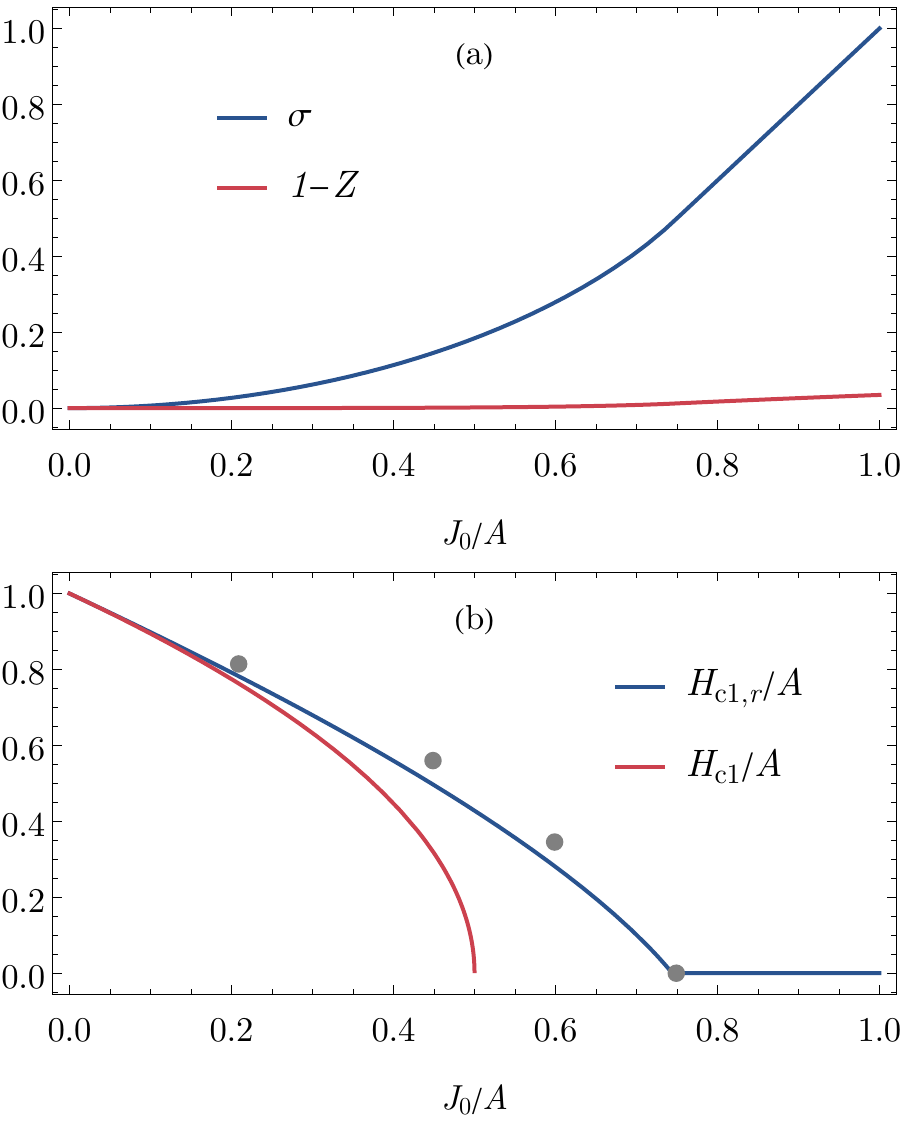}
\caption{
(a) Exchange and quasiparticle residue renormalizations of the triplet modes 
at $T=0$ in the quantum paramagnetic regime at the end of the flow 
for $\Lambda = 1$,
as function of the inter-dimer exchange coupling 
$ J^\bot_{ T , \bm{k} = 0 } = J^\parallel_{ T , \bm{k} = 0 } = J_0 $.
(b) Renormalized lower quantum critical field 
$ H_{ c 1 , r } = Z_{ \Lambda = 1 }^\bot 
\sqrt{ A^2 \left( 1 + \sigma_{ \Lambda = 1 }^\bot \right) - 2 A J^\bot_{ T , \bm{k} = 0 }   } $
as function of the inter-dimer exchange coupling 
$ J^\bot_{ T , \bm{k} =0 } = J^\parallel_{ T , \bm{k} =0 } = J_0 $,
with the mean-field  result  $H_{ c 1 }$ [Eq.~\eqref{eq:hc1}] for comparison.
The gray circles are the quantum Monte Carlo results 
of Ref.~[\onlinecite{Nohadani2004}].
}
\label{fig:T0}
\end{figure}

At larger values of the inter-dimer exchange couplings
and low temperatures,
quantum fluctuations can also become important in the quantum paramagnetic phase.
In particular,
quantum fluctuations  renormalize  the lower critical field $H_{c1}$ 
given in Eq.~\eqref{eq:hc1} \cite{Nohadani2004}.
To investigate this effect,
we need only consider the spin FRG flow equations 
for the staggered $2$-point vertex functions
at zero temperature, 
which are given by
\begin{subequations} \label{eq:T=0_flow}
\begin{align}
&
\partial_\Lambda \Gamma_{ \Lambda , T T }^{ + - } ( - K , K )
\nonumber\\
= {} &
\int_Q \Gamma_{ \Lambda , T T T T }^{ + + - - } ( - K , -Q , Q , K ) 
\dot{ G }_ {\Lambda , T }^\bot ( Q )
\nonumber\\
&
+ \frac{ 1 }{ 2 }
\int_Q \Gamma_{ \Lambda , T T T T }^{ + - z z } ( - K , K , -Q , Q ) 
\dot{ G }_ {\Lambda , T }^\parallel ( Q ) ,
\\
&
\partial_\Lambda \Gamma_{ \Lambda , T T }^{ z z } ( - K , K )
\nonumber\\
= {} &
\int_Q \Gamma_{ \Lambda , T T T T }^{ + - z z } ( - Q , Q , -K , K ) 
\dot{ G }_ {\Lambda , T }^\bot ( Q )
\nonumber\\
&
+ \frac{ 1 }{ 2 }
\int_Q \Gamma_{ \Lambda , T T T T }^{ z z z z } ( - Q , Q , -K , K ) 
\dot{ G }_ {\Lambda , T }^\parallel ( Q ) ,
\end{align}
\end{subequations}
and shown graphically in Fig.~\ref{fig:gamma2}.
Since we aim to describe the quantum paramagnetic phase at $T=0$,
the magnetization vanishes for all values of the deformation parameter $\Lambda$,
$ M_\Lambda = 0 $,
so that there are no tadpole corrections to the vertex functions.
Neglecting higher-order loop corrections as in Sec.~\ref{sec:thermal},
we may approximate the $4$-point vertices in the above flow equations \eqref{eq:T=0_flow}
by their initial values which
reflect the non-trivial quantum dynamics of the staggered spin of an 
isolated dimer. As shown in Appendix \ref{app:dimer},  
for the frequency-arguments needed in the
flow equations \eqref{eq:T=0_flow}, the initial values of 
the three different $4$-point vertices associated with 
the staggered spin are
\begin{subequations}
\begin{align}
\Gamma_{ 0 , T T T T }^{ + + - - } ( - i\omega , - i\nu , i\nu , i\omega ) = {} 
&
\frac{ 1 }{ A^3 }
\left[ A^2 + ( H - i\omega ) ( H - i\nu )  \right]
\nonumber\\
& \times
\left[ A^2 - \left( H - \frac{ i\omega + i\nu }{ 2 } \right)^2 \right] ,
\\
\Gamma_{ 0 , T T T T }^{ + - z z } ( - i\omega , i\omega , - i\nu , i\nu ) = {} 
&
\frac{ 1 }{ 2 A^3 }
\left[ A^4 - ( H - i\omega )^2 ( i\nu )^2  \right] ,
\\
\Gamma_{ 0 , T T T T }^{ z z z z } ( - i\omega , i\omega , - i\nu , i\nu ) = {} 
&
\frac{ 1 }{ 2 A^3 }
\left[ 3 A^4 - ( i\omega )^2 ( i\nu )^2 \right.
\nonumber\\
& \phantom{ \frac{ 1 }{ 2 A^3 } } \left.
- A^2  (i\omega )^2 - A^2 ( i\nu )^2  \right] .
\end{align}
\end{subequations}
Then it turns out that the staggered $2$-point vertex functions 
can be parametrized as
\begin{subequations}
\begin{align}
\Gamma_{ \Lambda , T T }^{ + - } ( - K , K )
= {} &
\frac{ A }{ 2 } \left( 1 + \sigma_\Lambda^\bot \right) 
- A \left( \frac{ H -i\omega }{ A Z_\Lambda^\bot } \right)^2 
+ J_{ T , \bm{k} }^\bot
, \\
\Gamma_{ \Lambda , T T }^{ z z } ( - K , K )
= {} &
\frac{ A }{ 2 } \left( 1 + \sigma_\Lambda^\parallel \right) 
- A \left( \frac{ i\omega }{ A Z_\Lambda^\parallel } \right)^2 
+ J_{  T , \bm{k} }^\parallel
,
\end{align}
\end{subequations}
where $\sigma_\Lambda^\alpha$ and $Z_\Lambda^\alpha$
are flowing renormalizations of exchange and quasiparticle residue respectively,
with initial conditions
$ \sigma_0^\alpha = 0 $
and
$ Z_0^\alpha = 1 $. The transverse and longitudinal components
of these couplings
satisfy the flow equations
\begin{widetext}
\begin{subequations} \label{eq:T0_transverse}
\begin{align}
\partial_\Lambda \sigma^\bot_\Lambda  
&=  
\left( Z_\Lambda^\bot \right)^2 \int_{ \bm{k} } 
\dot{ J }^\bot_{ \Lambda , T , \bm{k} } 
\frac{ \partial }{ \partial J^\bot_{ \Lambda , T , \bm{k} } }
\frac{ 4 A^2 - \epsilon_\Lambda^2 ( J^\bot_{ \Lambda , T , \bm{k} } ) 
}{ 2 A \epsilon_\Lambda ( J^\bot_{ \Lambda , T , \bm{k} } ) } 
+ 
\left( Z_\Lambda^\parallel \right)^2 \int_{ \bm{k} } 
\dot{ J }^\parallel_{ \Lambda , T , \bm{k} } 
\frac{ \partial }{ \partial J^\parallel_{ \Lambda , T , \bm{k} } }
\frac{ A }{ 2 \epsilon_\Lambda ( J^\parallel_{ \Lambda , T , \bm{k} } ) }  ,
\\
\frac{ \partial_\Lambda Z^\bot_\Lambda }{ \left( Z_\Lambda^\bot \right)^3 } 
&=
-
\left( Z_\Lambda^\bot \right)^2 \int_{ \bm{k} } 
\dot{ J }^\bot_{ \Lambda , T , \bm{k} }
\frac{ \partial }{ \partial J^\bot_{ \Lambda , T , \bm{k} } }
\frac{ A^2 + 2 \epsilon_\Lambda^2 ( J^\bot_{ \Lambda , T , \bm{k} } ) 
}{ 4 A \epsilon_\Lambda ( J^\bot_{ \Lambda , T , \bm{k} } ) } 
-
\left( Z_\Lambda^\parallel \right)^2 \int_{ \bm{k} } 
\dot{ J }^\parallel_{ \Lambda , T , \bm{k} }
\frac{ \partial }{ \partial J^\parallel_{ \Lambda , T , \bm{k} } }
\frac{ \epsilon_\Lambda ( J^\parallel_{ \Lambda , T , \bm{k} } ) }{ 4 A }  ,
\end{align}
\end{subequations}
and
\begin{subequations} \label{eq:T0_parallel}
\begin{align}
\partial_\Lambda \sigma^\parallel_\Lambda  
&=  
\left( Z_\Lambda^\bot \right)^2 \int_{ \bm{k} } 
\dot{ J }^\bot_{ \Lambda , T , \bm{k} } 
\frac{ \partial }{ \partial J^\bot_{ \Lambda , T , \bm{k} } }
\frac{ A 
}{ \epsilon_\Lambda ( J^\bot_{ \Lambda , T , \bm{k} } ) } 
+
\left( Z_\Lambda^\parallel \right)^2 \int_{ \bm{k} } 
\dot{ J }^\parallel_{ \Lambda , T , \bm{k} }
\frac{ \partial }{ \partial J^\parallel_{ \Lambda , T , \bm{k} } }
\frac{ 3 A^2 - \epsilon_\Lambda^2 ( J^\parallel_{ \Lambda , T , \bm{k} } ) 
}{ 2 A \epsilon_\Lambda ( J^\parallel_{ \Lambda , T , \bm{k} } ) } ,
\\
\frac{ \partial_\Lambda Z^\parallel_\Lambda }{ \left( Z_\Lambda^\parallel \right)^3 } 
&=
-
\left( Z_\Lambda^\bot \right)^2 \int_{ \bm{k} } 
\dot{ J }^\bot_{ \Lambda , T , \bm{k} } 
\frac{ \partial }{ \partial J^\bot_{ \Lambda , T , \bm{k} } }
\frac{ \epsilon_\Lambda ( J^\bot_{ \Lambda , T , \bm{k} } ) 
}{ 2 A } 
-
\left( Z_\Lambda^\parallel \right)^2 \int_{ \bm{k} } 
\dot{ J }^\parallel_{ \Lambda , T , \bm{k} } 
\frac{ \partial }{ \partial J^\parallel_{ \Lambda , T , \bm{k} } }
\frac{ A^2 + \epsilon_\Lambda^2 ( J^\parallel_{ \Lambda , T , \bm{k} } ) 
}{ 4 A \epsilon_\Lambda ( J^\parallel_{ \Lambda , T , \bm{k} } ) } ,
\end{align}
\end{subequations}
\end{widetext}
where 
\begin{equation}
\epsilon^\alpha_\Lambda ( J )
=
Z_\Lambda^\alpha 
\sqrt{ A^2 \left( 1 + \sigma_\Lambda^\alpha \right) + 2 A J  }
\end{equation}
are the flowing dispersion relations of the triplet modes
at zero temperature.
These flow equations are again of the form of Eq.~\eqref{eq:Litim_loop}
and  with the Litim regulator \eqref{eq:Litim}
 reduce to ordinary differential equations.
Note especially that for isotropic flowing inter-dimer exchange couplings
$J_{ \Lambda, T , \bm{k} }^\bot = J_{ \Lambda, T , \bm{k} }^\parallel$,
the respective flow equations \eqref{eq:T0_transverse} and \eqref{eq:T0_parallel}
for the transverse and longitudinal renormalizations are identical.
Hence, 
$\sigma_\Lambda^\bot = \sigma_\Lambda^\parallel = \sigma_\Lambda$ and
$Z_\Lambda^\bot = Z_\Lambda^\parallel = Z_\Lambda$ in this case.
The resulting renormalization of the triplet modes
and the lower quantum critical field
at the end of the flow are shown in Fig.~\ref{fig:T0}
for inter-dimer exchange couplings
$J_{ T , \bm{k} = 0 }^\bot = J_{ T , \bm{k} = 0 }^\parallel = J_0$
as function of $J_0$.
It can be seen that
for inter-dimer exchange couplings 
$ J_0 \gtrsim 0.2 \, A $,
quantum fluctuations lead to a significant renormalization
of the triplet dispersions and consequently of the lower quantum critical field.
On the other hand,
the triplet modes remain well defined,
$Z \approx 1$,
even for larger inter-dimer exchange couplings.
Note especially that
our result for the dependence of the lower quantum critical field 
on the inter-dimer exchange agrees both qualitatively and quantitatively 
rather well with the quantum Monte Carlo simulation results
of Ref.~[\onlinecite{Nohadani2004}]
at \emph{all} values of the inter-dimer exchange.

\section{Summary and outlook}

\label{sec:discussion}

The present work has established the applicability and power of the 
recently developed spin FRG formalism  \cite{Krieg2019,Tarasevych2018,Goll2019,Goll2020,Tarasevych2021,Tarasevych2022} 
for dimerized quantum spin systems.
Using a deformation  scheme where the spin-correlation functions of isolated
dimers define the initial conditions for the FRG flow,
we have shown that even relatively simple truncations of the flow equations
yield quantitatively accurate  results for the spectrum and thermodynamics 
in the entire quantum paramagnetic, ferromagnetic,
and thermally disordered phases.
In particular,
we have found that retaining the tadople diagrams to all orders
generates a self-consistent mean-field correction to the magnetic field,
which acts as a chemical potential for the triplet excitations.
With this key ingredient,
we have solved the flow equations for the free energy and the magnetization
in a one-loop truncation. 
The critical softening of the lowest triplet mode has then allowed us 
to determine the critical magnetic field for the phase transition to the
antiferromagnetic $XY$ phase at \emph{all} temperatures.
At low enough temperatures,
our flow equations have furthermore recovered the established critical exponents 
that are expected for the two BEC quantum critical points. 
Lastly,
we have demonstrated 
that we can also include quantum fluctuations in the quantum paramagnetic phase
by deriving and solving flow equations that describe 
the renormalization of the triplet modes,
and thereby also of the lower quantum critical field,
at zero temperature.

An alternative functional renormalization group approach to quantum spin systems is based on the representation of the spin-operators in terms of Abrikosov 
pseudofermions \cite{Reuther10,Reuther11,Reuther11a,Buessen16,Thoenniss20,Kiese20,Ritter22} or Majorana fermions \cite{Niggemann21} and the numerical solution of the resulting 
truncated fermionic FRG flow equations. Apparently, so far this
pseudofermion FRG approach has not been applied  to dimerized spin systems.
Our spin FRG suggests that this would require a proper parametrization of the quantum dynamics encoded in  $4$-spin correlations of an isolated dimer,
which seems to be rather difficult within the pseudofermion FRG.

Finally, let us point out that
this work can be extended in several directions:
On the one side,
one could investigate also the antiferromagnetically ordered phase,
in principle in arbitrary dimensions. 
On the other side,
it would be interesting to study the interactions and damping of the triplet modes,
in particular in the quantum critical regimes 
that are already accessible with the present setup of the spin FRG. 
Finally,
it might be interesting to consider dimerized spin systems 
on more complicated lattices,
such that quantitative comparisons with experiments come within reach.

\begin{acknowledgments}
This work was financially supported by the Deutsche
Forschungsgemeinschaft (DFG, German Research Foundation) 
through Project No. KO/1442/10-1.
We appreciate fruitful discussions with 
Bernd Wolf and Michael Lang during the early stages of this work.
\end{acknowledgments}

\appendix

\setcounter{equation}{0}

\renewcommand{\theequation}{A\arabic{equation}}

\renewcommand{\appendixname}{APPENDIX}

\renewcommand{\thesection}{\Alph{section}}

\section{TIME-ORDERED CORRELATION FUNCTIONS OF AN
ISOLATED DIMER} 

\label{app:dimer}

\renewcommand{\theequation}{A\arabic{equation}}

\begin{figure}[tb]
\centering
\includegraphics[width=\linewidth]{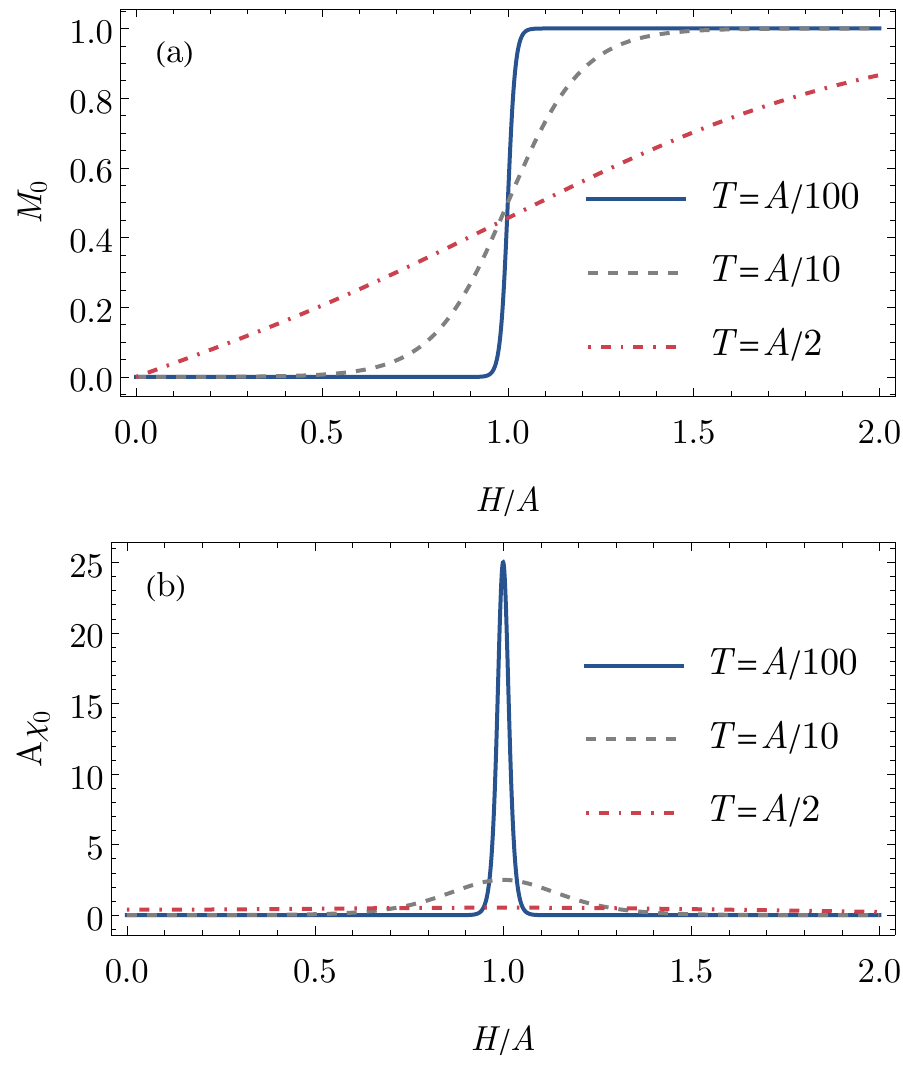}
\caption{
(a) Magnetization \eqref{eq:M0} and
(b) susceptibility \eqref{eq:chi0} of the isolated dimer,
as function of applied magnetic field for different temperatures.
}
\label{fig:MChi0}
\end{figure}

The isolated dimer consisting of two $S=1/2$ spins 
is central to our formulation of the spin FRG 
for dimerized quantum spin systems,
because its imaginary time-ordered correlation functions define the initial condition 
of the FRG flow at 
$ J_{ \Lambda = 0 , s , i j }^\alpha =  0$.
Therefore we devote this Appendix to a short overview 
of the salient features of the isolated dimer,
and the computation of the relevant correlation functions.
The Hamiltonian of an isolated  dimer reads
\begin{equation} \label{eq:h_dimer0}
h = \frac{ A }{ 2 } \bm{ S }^2 - H S^z ;
\end{equation}
see Eq.~\eqref{eq:h_dimer}.
Here, $\bm{ S } = \bm{s}_1 + \bm{s}_2$ denotes the total spin operator of the  dimer
as defined in Eq.~\eqref{eq:ST}, where $\bm{s}_1$ and $\bm{s}_2$ are two independent 
spin-$1/2$ operators.
In the singlet-triplet eigenbasis 
of the dimer Hamiltonian \eqref{eq:h_dimer0}
that is discussed in Sec.~\ref{sec:dimer},
the staggered and total spin operators explicitly read
\begin{subequations} \label{eq:ST_matrix}
\begin{align}
	&T^{+} =\begin{pmatrix}
		0& 0& 0& 1\\
		-1& 0& 0& 0\\
		0& 0& 0& 0\\
		0& 0& 0& 0
	\end{pmatrix}=(T^-)^\dagger,\\
	&T^{z} = \begin{pmatrix} 
		0& 0& 1& 0\\
		0& 0& 0& 0\\
		1& 0& 0& 0\\
		0& 0& 0& 0
	\end{pmatrix},\\
	&S^{+} =\begin{pmatrix} 
		0& 0& 0& 0\\
		0& 0& 1& 0\\
		0& 0& 0& 1\\
		0& 0& 0& 0
	\end{pmatrix} =(S^-)^\dagger,\\
	&S^{z} =\begin{pmatrix} 
		0& 0& 0& 0\\
		0& 1& 0& 0\\
		0& 0& 0& 0\\
		0& 0& 0& -1
	\end{pmatrix} ,
\end{align}
\end{subequations}
where we ordered the states in ascending order corresponding to their eigenenergies,
assuming that the singlet has the lowest energy.
The partition function of the isolated dimer is
\begin{equation}
Z 
= \frac{ 1 }{ p^{ s } }
= 1 + e^{ - \beta E^{+} } 
+ e^{ - \beta E^{0} } + e^{ - \beta E^{-}} ,
\end{equation}
where the eigenenergies are given in Eq.~(\ref{eq:dimer_energies}).
From the corresponding free energy,
\begin{equation}
	f_0 = - \frac{ 1 }{ \beta } \ln  Z  ,
\end{equation}
we obtain the magnetization,
\begin{equation} \label{eq:M0}
	M_0 = - \frac{ \partial f_0 }{ \partial H } = p^{+} - p^{-} ,
\end{equation}
and the static longitudinal susceptibility,
\begin{equation} \label{eq:chi0}
\chi_0
= \frac{ \partial M_0 }{ \partial H } 
= \beta \left( p^{ + } +  p^{ - } - M_0^2 \right) .
\end{equation}
The Boltzmann weights are given in Eq.~(\ref{eq:Boltzmann}).
To gain some intuitive understanding 
of the behavior of the isolated dimer,
we plot in Fig.~\ref{fig:MChi0}   the magnetization \eqref{eq:M0} 
and static susceptibility \eqref{eq:chi0}
as functions of the applied magnetic field for various temperatures.
Because of the time reversal symmetry
of the dimer Hamiltonian \eqref{eq:h_dimer0},
we can focus on $ H > 0 $ without loss of generality.
At zero temperature and for magnetic fields $ H $ 
smaller than the intra-dimer exchange $ A $, 
the dimer is in the singlet state, 
$ p^{s} = 1 $. 
Precisely at $ H = A $, 
the dimer undergoes a field induced quantum phase transition 
into the fully polarized $+$ triplet state 
with $ p^{+} = 1 $. 
Thus, at zero temperature the magnetization is a simple step function, 
$ M_0(T=0) = \Theta( H - A ) $.
At finite temperatures, on the other hand,
all dimer states are thermally occupied 
according to their respective Boltzmann factors \eqref{eq:Boltzmann},
implying a smooth magnetization curve.
Saturation is no longer reached 
once the temperature is sufficient 
to excite non-magnetic states. 
Further increasing the temperature results in 
the magnetization becoming more linear, 
with a slope proportional to the inverse temperature. 
Correspondingly,
the susceptibility exhibits a single $\delta$-like peak 
at the critical field $ H = A $,
which widens and becomes field independent in leading order 
towards high temperatures.

For the initial conditions of the vertex expansion,
we require the imaginary-time ordered connected $n$-point
correlation function of the staggered and total dimer spin as well.
They are defined as
\begin{widetext}
 \begin{align} \label{eq:G0_connected}
\delta( \omega_1 + \ldots + \omega_n ) 
G_{ 0 ,
\underbrace{ \scriptstyle S \ldots S }_{ m } 
\underbrace{ \scriptstyle T \ldots T }_{ n-m } }
^{ \alpha_1 \ldots \alpha_n }
( i\omega_1 , \ldots , i\omega_n )
= {} &
\int_{ 0 }^{ \beta } d \tau_1 \ldots \int_{ 0 }^{ \beta } d\tau_{ n } 
e^{ i ( \omega_1 \tau_1 + \ldots + \omega_{ n } \tau_{ n } ) } 
\nonumber\\
& 
\times
\braket{ {\cal T } 
S^{ \alpha_1 }( \tau_1 ) \ldots S^{ \alpha_m }( \tau_m ) 
T^{ \alpha_{ m + 1 } }( \tau_{ m + 1 } )\ldots 
T^{ \alpha_{ n } }( \tau_{ n } ) }_{\rm connected } .
\end{align}
\end{widetext}
Here, ${\cal{T}}$ denotes time-ordering in imaginary time, and
the imaginary time dependence of the operators 
is in the Heisenberg picture.
We also used the energy conservation to factor out the frequency-$\delta$,
which is defined as $\delta(\omega) = \beta \delta_{ \omega , 0 }$.
In frequency space, 
the connected spin correlation functions \eqref{eq:G0_connected} 
can be calculated
efficiently using their spectral representations. 
Since the eigenenergies \eqref{eq:dimer_energies} 
of the Hamiltonian \eqref{eq:h_dimer0}
as well as the matrix representations \eqref{eq:ST_matrix} 
of the spin operators are known,
one can carry out the required Fourier transformation explicitly
\cite{Izyumov1988}.
The $2$-point functions are given by
\begin{subequations} \label{eq:2-point_dimer}
\begin{align}
G_{ 0 , T T }^{ + - } ( i\omega , - i\omega)
&= G_{ 0 , T }^\bot ( i\omega )
= \sum_{ r = \pm } \frac{ p^{ s } - p^{ r } }{ A - r \left( H - i\omega \right) } ,
\\
G_{ 0 , T T }^{ z z } ( i\omega , - i\omega)
&= G_{ 0 , T }^\parallel ( i\omega )
= \frac{ 2 A \left( p^{ s } - p^{ 0 } \right) }{ A^2 + \omega^2 } ,
\\
G_{ 0 , S S }^{ + - } ( i\omega , - i\omega)
&= G_{ 0 , S }^\bot ( i\omega )
=  \frac{ M_0 }{ H - i\omega } ,
\\
G_{ 0 , S S }^{ z z } ( i\omega , - i\omega)
&= G_{ 0 , S }^\parallel ( i\omega )
= \delta_{ \omega , 0 } \chi_0 .
\label{eq:Gzz0}
\end{align}
\end{subequations}
Note that the longitudinal total spin has no dynamics,
$G_{ 0 , S }^\parallel ( i\omega ) \propto \delta_{ \omega , 0 }$,
reflecting the $U(1)$ spin-rotational symmetry of the dimer Hamiltonian \eqref{eq:h_dimer0}
around the direction of the magnetic field. 
The finite longitudinal and mixed transverse-longitudinal $3$-point functions are
\begin{widetext}
\begin{subequations}
\begin{align}
G_{ 0 , T T S }^{ z z z } ( i\omega_1 , i\omega_2 , i\omega_3 )
= {} &
- \delta( \omega_3 ) M_0 G_{ 0 , T }^\parallel ( i\omega_3 ) ,
\\
G_{ 0 , S S S }^{ z z z } ( i\omega_1 , i\omega_2 , i\omega_3 )
= {} &
\delta( \omega_2 ) \delta( \omega_3 ) M_0
\left( 1 - 3 \beta^{ -1 } \chi_0 - M_0^2 \right)
=
\delta_{ \omega_2 , 0 } \delta_{ \omega_3 , 0 } \partial_H \chi_0 ,
\end{align}
\end{subequations}
and
\begin{subequations}
\begin{align}
G_{ 0 , T T S }^{ + - z } ( i\omega_1 , i\omega_2 , i\omega_3 )
= {} &
\frac{ G_{ 0 , T }^\bot ( i\omega_1 ) - G_{ 0, T }^\bot ( - i\omega_2 ) 
}{ i\omega_3 }
- \delta( \omega_3 ) \left[
\sum_{ r = \pm } \frac{ r p^{ r } }{ A - r \left( H - i\omega_1 \right) }
+ M_0 G_{ 0 , T }^\bot ( i\omega_1 )
\right] ,
\\
G_{ 0 , T S T }^{ + - z } ( i\omega_1 , i\omega_2 , i\omega_3 )
= {} &
\frac{ G_{ 0 , S }^\bot ( i\omega_1 ) }{ M_0 } \left[
G_{ 0 , T }^\parallel ( i\omega_3 ) - G_{ 0 , T }^\bot ( i\omega_1 )
\right] ,
\\
G_{ 0 , S T T }^{ + - z } ( i\omega_1 , i\omega_2 , i\omega_3 )
= {} &
\frac{ G_{ 0 , S }^\bot ( i\omega_1 ) }{ M_0 } \left[
G_{ 0 , T }^\parallel ( i\omega_3 ) - G_{ 0 , T }^\bot ( - i\omega_2 )
\right] ,
\\
G_{ 0 , S S S }^{ + - z } ( i\omega_1 , i\omega_2 , i\omega_3 )
= {} &
\frac{ G_{ 0 , S }^\bot ( i\omega_1 ) }{ M_0 } \left[
- G_{ 0 , S }^\bot ( - i\omega_2 ) 
+ M_0 G_{ 0 , S }^\parallel ( i\omega_3 ) 
\right] .
\end{align}
\end{subequations}
For the calculation of the quantum fluctuations in Sec.~\ref{sec:quantum},
we also require the initial conditions of the staggered $4$-point vertices,
which are determined by the staggered $4$-point correlation functions
of the isolated dimer via the tree expansion;
see Eq.~\eqref{eq:tree} and 
Refs.~[\onlinecite{Kopietz2010,Krieg2019,Goll2020}].
After some tedious calculations  we find that the different components of the
staggered $4$-spin correlation functions of an isolated dimer  are given by the following expressions,
\begin{subequations}
\begin{align}
&
G_{ 0 , T T T T }^{ + + - - } ( i\omega_1 , \ldots , i\omega_4 )
\nonumber\\
= {} &
\frac{
\left[ 2 \left( A + H \right) + i\omega_3 + i\omega_4 \right]
\left[ 2 \left( A - H \right) - i\omega_3 - i\omega_4 \right]
}{
2 H + i\omega_3 + i\omega_4
}
\nonumber\\
& 
\times
\sum_{ r = \pm } 
\frac{ r \left( p^r - p^s \right)
}{
\left[ A - r \left( H - i\omega_1 \right) \right] 
\left[ A - r \left( H - i\omega_2 \right) \right]
\left[ A - r \left( H + i\omega_3 \right) \right] 
\left[ A - r \left( H + i\omega_4 \right) \right]
}
\nonumber\\[.2cm]
&
+ \left[ 
\delta\left( \omega_1 + \omega_3 \right) +
\delta\left( \omega_1 + \omega_4 \right)
\right]
\nonumber\\
&
\times
\left\{
\sum_{ r = \pm }
\frac{ p^{ r  } }{ 
\left[ A - r \left( H - i\omega_1 \right) \right] 
\left[ A - r \left( H - i\omega_2 \right) \right] }
+ \frac{ 4 A^2 p^{  s  } }{ 
\left[ A^2 - \left( H - i\omega_1 \right)^2 \right] 
\left[ A^2 - \left( H - i\omega_2 \right)^2 \right] }
- G_{ 0 , T }^\bot ( i\omega_1 ) G_{ 0 , T }^\bot ( i\omega_2 )
\right\} ,
\\[.25cm]
&
G_{ 0 , T T T T }^{ + - z z } ( i\omega_1 , \ldots , i\omega_4 )
\nonumber\\
= {} &
\sum_{ r = \pm } \frac{ r p^{ r } }{
\left[ A - r \left( H - i\omega_1 \right) \right]
\left[ A - r \left( H + i\omega_2 \right) \right]
}
\left(  
\frac{ 1 }{ H + i\omega_2 + i\omega_3 } +
\frac{ 1 }{ H + i\omega_2 + i\omega_4 }
\right)
\nonumber\\
&
- p^{ s } \sum_{ r = \pm } \left\{
\frac{ 1 }{ \left[ A +r \left( H - i\omega_1 \right) \right] \left( A - r i\omega_3 \right) }
\left(  
\frac{ 1 }{ A + r \left(H + i\omega_2 \right) } +
\frac{ 1 }{ A + r i\omega_4 }
\right)
+ \left( i\omega_3 \leftrightarrow i\omega_4 \right)
\right\}
\nonumber\\
&
- \frac{ 2 A p^{ 0 } \left( i\omega_3 - i\omega_4 \right)^2 }{
\left( H + i\omega_2 + i\omega_3 \right)
\left( H + i\omega_2 + i\omega_4 \right)
\left( A^2 + \omega_3^2 \right)
\left( A^2 + \omega_4^2 \right)
}
\nonumber\\[.2cm]
&
+ \delta\left( \omega_1 + \omega_2 \right)
\left\{
\frac{ 4 A^2 p^{ s } }{ 
\left[ A^2 - \left( H - i\omega_1 \right)^2 \right]
\left( A^2 + \omega_3^2 \right)
}
- G_{ 0 , T }^\bot ( i\omega_1 ) G_{ 0 , T }^\parallel ( i\omega_3 )
\right\} ,
\\[.45cm]
&
G_{ 0 , T T T T }^{ z z z z } ( i\omega_1 , \ldots , i\omega_4 )
\nonumber\\
= {} &
\frac{ 4 A^2 \left( p^{ 0 } - p^{ s } \right) 
\left[
6 A^4 - 2 \omega_1 \omega_2 \omega_3 \omega_4 
+ A^2 \left( \omega_1^2 + \omega_2^2 + \omega_3^2 + \omega_4^2 \right)
\right]
}{
\left( A^2 + \omega_1^2 \right)
\left( A^2 + \omega_2^2 \right)
\left( A^2 + \omega_3^2 \right)
\left( A^2 + \omega_4^2 \right)
}
\nonumber\\[.2cm]
&
+ \left\{
\delta\left( \omega_1 + \omega_2 \right) \left[
\frac{ 4 A^2 \left( p^{ 0 } + p^{ s } \right) 
}{
\left( A^2 + \omega_1^2 \right)
\left( A^2 + \omega_3^2 \right)
}
- G_{ 0 , T }^\parallel ( i\omega_1 ) G_{ 0 , T }^\parallel ( i\omega_3 )
\right]
+ \left( i\omega_2 \leftrightarrow i\omega_3 \right)
+ \left( i\omega_2 \leftrightarrow i\omega_4 \right)
\right\} .
\end{align}
\end{subequations}
\end{widetext}

\section{DETAILS OF THE SPIN FRG FORMALISM} 

\label{app:SFRG}

\renewcommand{\theequation}{B\arabic{equation}}

In order to set up the spin FRG for our dimerized spin system \eqref{eq:H_spins},
it is convenient to introduce the following  compact notation:
We collect all field labels into a collective label $x \equiv  ( \alpha a i \tau )$,
where $\alpha = x , y , z$ labels the Cartesian component,
$a = S , T$ the field flavor, 
$i = 1, \ldots , N$ the dimer,
and $\tau$ is the imaginary time.
We then define the collection $I_x$ of total and staggered spin operators such that 
\begin{equation}
\left( \begin{matrix}
I_{ S , i }^x ( \tau ) \\
I_{ S , i }^y ( \tau ) \\
I_{ S , i }^z ( \tau ) \\
I_{ T , i }^x ( \tau ) \\
I_{ T , i }^y ( \tau ) \\
I_{ T , i }^z ( \tau )
\end{matrix} \right)
=
\left( \begin{matrix}
S_{ i }^x ( \tau ) \\
S_{ i }^y ( \tau ) \\
S_{ i }^z ( \tau ) \\
T_{ i }^x ( \tau ) \\
T_{ i }^y ( \tau ) \\
T_{ i }^z ( \tau )
\end{matrix} \right) .
\end{equation}
In the space of this collective label,
the (deformed) inter-dimer exchange matrix is given by
\begin{align}
\left( {\bf J}_\Lambda \right)_{ x x' } 
= {} &
\delta^{ \alpha \alpha' } \delta_{ a a' } \delta( \tau - \tau' )
\nonumber\\
& \times \left[
\left( 1 
- \delta^{ \alpha z } \right) J_{ \Lambda , a , i i' }^\bot 
+ \delta^{ \alpha z } J_{ \Lambda , a , i i' }^\parallel 
\right] .
\end{align}
With this compact notation,
the generating functional of connected spin correlation functions can be written as
\begin{equation} \label{eq:generating_functional_connected}
{\cal G}_\Lambda [ h ] 
=
\ln {\rm Tr} \left[
e^{ - \beta {\cal H}_0 } {\cal T} e^{ \int_x h_x I_x - \frac{ 1 }{ 2 } \int_{ x x' } I_x \left( {\bf J}_\Lambda \right)_{ x x' } I_{ x' } }
\right] ,
\end{equation}
where $h_x$ is the source field conjugate to the operator $I_x$,
the integration symbol is $\int_x = \sum_\alpha \sum_a \sum_i \int_0^\beta d\tau$,
and ${\cal T}$ denotes imaginary-time ordering of everything to its right.
All spin operators $I_x$ are taken to be in the imaginary-time Heisenberg picture with respect to the Hamiltonian ${\cal H}_0$ of decoupled dimers
given in Eq.~\eqref{eq:H_0}.
Ideally, we would like to work exclusively with one-line irreducible vertices \cite{Kopietz2010,Krieg2019}. 
Their generating functional is the Legendre transformation of the generating functional \eqref{eq:generating_functional_connected} of connected correlation functions. However, 
because the dimer Hamiltonian \eqref{eq:H_0} possesses $U(1)$ spin-rotational symmetry around the $z$ axis, 
such a Legendre transformation is not well-defined when we turn off the inter-dimer exchange ${\bf J}_\Lambda$ at the initial scale of the RG flow \cite{Goll2019}. 
Ultimately, the reason for this is that $S_i^z$ is conserved for each dimer and hence has no dynamics [see Eq.~\eqref{eq:Gzz0}], 
which makes it impossible to express the conjugate source via the respective average field. 
To circumvent this issue, it is convenient \cite{Goll2019} 
to work with a hybrid functional
\begin{equation} \label{eq:hybrid}
{\cal F}_\Lambda [ h ] 
=
{\cal G}_\Lambda \left[ ( {\bf 1} - {\bf P} ) h - {\bf P} {\bf J}_\Lambda h \right] 
- \frac{ 1 }{ 2 } \int_{ x x' } h_x \left( {\bf P} {\bf J}_\Lambda \right)_{ x x' } h_{ x' } ,
\end{equation}
where the matrix ${\bf P} = {\bf P}^2$ with components 
\begin{equation}
\left( {\bf P} \right)_{ x x' } 
=
\delta^{ \alpha \alpha' } \delta_{ a a' } \delta_{ i i'} \delta( \tau - \tau' ) \delta^{ \alpha z } \delta_{ a S }
\end{equation}
projects onto the longitudinal magnetization subspace.
The functional \eqref{eq:hybrid} generates connected correlation functions of the staggered and the transverse dimer spin 
and amputated connected longitudinal dimer spin correlation functions. 
The associated generating functional of one-line irreducible vertices is then well-defined at the initial scale;
it is explicitly given by
\begin{align} 
\Gamma_\Lambda [ \varphi ] 
= {} &
\int_x h_x \left( \phi_{ \Lambda , x } + \varphi_x \right) - {\cal F}_\Lambda [ h ] 
\nonumber\\
&
- \frac{ 1 }{ 2 } \int_{ x x' } 
\varphi_x \left( {\bf R}_\Lambda \right)_{ x x' } \varphi_{ x' } ,
\label{eq:Gamma_functional}
\end{align}
where 
\begin{equation}
{\bf R}_\Lambda = 
\left( {\bf 1} - {\bf P} \right) \left( {\bf J}_\Lambda - {\bf J} \right) \left( {\bf 1} - {\bf P} \right) 
+ {\bf P} \left( - {\bf J}_\Lambda^{ -1 } + {\bf J}^{ -1 } \right) {\bf P} 
\end{equation}
is the regulator matrix,
and the source fields $h_x [ \varphi ]$ are determined by inversion of 
\begin{equation}
\frac{ \delta {\cal F}_\Lambda [ h ] }{ \delta h_x } = \phi_{ \Lambda , x } + \varphi_x ,
\end{equation}
with the vacuum expectation values 
\begin{equation}
\left. \frac{ \delta {\cal F}_\Lambda [ h ] }{ \delta h_x } \right|_{ h = 0} = \phi_{ \Lambda , x } .
\end{equation}
Note that compared to Refs.~\cite{Krieg2019,Goll2019,Goll2020},
we use a slightly different regulator subtraction
with the fluctuating fields $\varphi_x$ 
instead of the full field $\phi_{ \Lambda , x } + \varphi_x$
in the second line of the generating functional \eqref{eq:Gamma_functional}.
This turns out to be more convenient in the presence 
of finite vacuum expectation values $\phi_{ \Lambda , x }$,
because $\phi_{ \Lambda , x }$ can then be chosen as 
the flowing field configuration 
that minimizes $ \Gamma_\Lambda [ \varphi ] $
for vanishing source fields $h_x$;
i.e.,
\begin{equation}
\left.
\frac{ \delta \Gamma_\Lambda [ \varphi ] }{ \delta \varphi_x }
\right|_{ \varphi = 0 } = 0 .
\end{equation}

By taking derivatives with respect to the deformation parameter $\Lambda$,
it can be shown that the generating functionals ${\cal G}_\Lambda [ h ]$, ${\cal F}_\Lambda [ h ]$,
and $\Gamma_\Lambda [ \varphi ]$ given in Eqs.~\eqref{eq:generating_functional_connected}, 
\eqref{eq:hybrid} and \eqref{eq:Gamma_functional} respectively satisfy exact flow equations
which determine the evolution of the associated correlation functions when the exchange interaction is gradually deformed.
Because the spin operators at different lattice sites commute,
these equations are formally identical to the FRG flow equations for bosons.
For an explicit derivation of these flow equations,
we refer to Refs.~\cite{Krieg2019,Goll2019}.
Here,
we only require the flow equation of the generating functional \eqref{eq:Gamma_functional} of irreducible vertex functions.
It has the form of a bosonic Wetterich equation~\cite{Wetterich1993}, 
\begin{align}
&
\partial_\Lambda \Gamma_\Lambda [ \varphi ] - \int_x \frac{ \delta \Gamma_\Lambda [ \varphi ] }{ \delta \varphi_x } \partial_\Lambda \phi_{ \Lambda , x }
\nonumber\\
= {} &
\frac{ 1 }{ 2 } {\rm Tr} \left\{ 
\left( \partial_\Lambda {\bf R}_\Lambda \right) \left[
\left( {\bf \Gamma}''_\Lambda [ \varphi ] + {\bf R}_\Lambda \right)^{ - 1 } + {\bf P} {\bf J}_\Lambda {\bf P}
\right]
\right\}
\nonumber\\
&
+ \frac{ 1 }{ 2 } \int_{ x x' } \phi_{ \Lambda , x } 
\left( \partial_\Lambda {\bf R}_\Lambda \right)_{ x x' } \phi_{ \Lambda , x' } 
\nonumber\\
&
+ \int_{ x x' } \varphi_x 
\partial_\Lambda \left[ 
\left( {\bf R}_\Lambda \right)_{ x x' } \phi_{ \Lambda , x' } \right]
,
\end{align}
where
\begin{equation}
\left( {\bf \Gamma}''_\Lambda [ \varphi ] \right)_{ x x' } = \frac{ \delta^{ 2 } \Gamma_\Lambda [ \varphi ] }{ \delta \varphi_x \delta \varphi_{ x' } } 
\end{equation}
denotes the matrix of second functional derivatives of the generating functional,
and the trace runs over the collective label $x$.

In the last step,
we have to specify the initial condition for the generating functional $\Gamma_\Lambda [\varphi]$.
In a vertex expansion scheme,
this initial condition can be obtained from the correlation functions of isolated dimers 
given in  Appendix~\ref{app:dimer} via the tree expansion \cite{Kopietz2010,Krieg2019,Goll2020}.
This amounts to expanding both sides of 
\begin{equation} \label{eq:tree}
\frac{ 
\delta^{ 2 } {\cal F}_\Lambda [ h ] 
}{ \delta h_x \delta h_{ x' } } 
=
\left( 
{\bf \Gamma}''_\Lambda [ \varphi ] + {\bf R}_\Lambda 
\right)^{ - 1 }_{ x x' } 
\end{equation}
in powers of the source fields $h_x$ and comparing coefficients at $\Lambda = 0$.

\end{document}